\documentclass[12pt]{article}
\usepackage{amssymb,amsmath,cite,mathdots,blkarray}

\def\beq{\begin{equation}}
\def\eeq{\end{equation}}
\def\bea{\begin{eqnarray*}}
\def\eea{\end{eqnarray*}}

\def\N0{ {\mathbb Z}_{+} }
\def\v0{ |d,r) }

\def\proof#1{{\bf Proof:} #1 $\blacksquare$ \medskip}
\def\id{\mbox{id}}

\def\tv{\tilde{V}}
\def\lv{\tilde{\Lambda}}

\newtheorem{lemma}{Lemma}

\newtheorem{thm}{Theorem}

\begin{document}

\begin{center}
{\large\bf Reduced matrix elements of the orthosymplectic Lie superalgebra}\\
~~\\

{\large Mark D. Gould and Phillip S. Isaac}\\
~~\\

School of Mathematics and Physics, The University of Queensland, St Lucia QLD 4072, Australia.
\end{center}

\begin{abstract}
We utilise characteristic
identities to construct eigenvalue formulae for invariants and reduced matrix elements corresponding
to irreducible representations of $osp(m|n)$. In presenting these results, we
further develop our programme of constructive representation theory via characteristic identities.
\end{abstract}

%


\section{Introduction} \label{sec1}

It is widely accepted that Lie superalgebras and their representations \cite{Kac1977,Kac1978} play an essential
role in the analysis, utilization and ultimate understanding of supersymmetry in physical systems.
Such structures first came about in the setting of elementary particle physics (see
\cite{FayFer1977} and references therein) but have since been applied in a variety
of fields, including high energy physics \cite{CHR2012,dLMR2012} and condensed matter physics
\cite{BGM2012}. To provide some archetypes, Lie superalgebras appear as symmetry algebras of quantum
many-body boson-fermion systems, particularly those relating to the shell model of nuclear physics
\cite{BalBarIac1981}, they occur in the context of integrability in string theory
\cite{HoIwTs2009,Beis2012}, and also supersymmetric Yang-Mills theory \cite{BeiStau2003,Min2012}. In
most cases, it is not only the algebraic structure which is important, but also the representations.

Various aspects of the representation theory of basic classical Lie superalgebras (in
the sense of Kac) have been developed over the years, but the most relevant to the current work are
those that present matrix element formulae for various types of representations,
particularly in the articles
\cite{TolIstSmi1986,Palev1987,Palev1989,StoiVan2010,Molev2011,GIW1,GIW2}, all of which are related
to $gl(m|n)$. 
Of these articles, \cite{GIW1,GIW2} exhibit work undertaken by the current authors in order to develop techniques that make use of
characteristic identities. Specifically, the first paper \cite{GIW1} constructed invariants and reduced matrix
elements associated with representations of $gl(m|n)$, and the second article made use of these
results to produce matrix element formulae for the type 1 unitary representations of $gl(m|n)$. In
\cite{GIW2}, we highlight the significance of unitary representations in the context of canonical subalgebra embeddings. In
particular, due to the inner product (positive definite sesquilinear form) inherited by the
subalgebra, complete reducibility of a unitary representation follows immediately down the
subalgebra chain.

By contrast, the representations of $osp(m|2k)$ are in general not unitary, particularly due to the fact that
there is not a positive definite sesquilinear form defined on the
irreducible modules. The one exceptional case is $osp(2|2k)$ (i.e. when $m=2$), which is precisely the Lie
superalgebra $C(k+1)$ in the notation of Kac's classification \cite{Kac1977}. The Lie superalgebra
$osp(2|2k)$ is in fact a type I basic classical Lie superalgebra \cite{Kac1978}, the set of which also
includes $gl(m|n)$. The type I basic classical Lie superalgebras are well known 
to be the only Lie superalgebras that give rise to unitary representations \cite{SNR1977}, and such
unitary representations have been studied and classified by Gould and Zhang in a series of papers
\cite{ZhaGou19902,ZhaGou1990,GouZha1990}. 

In this paper, we consider the canonical subalgebra chain
\begin{equation}
osp(m+1|2k)\supset osp(m|2k) \supset \cdots \supset osp(2|2k)\supset osp (1|2k) \cdots
\label{subchain}
\end{equation}
Since the representations are not unitary in general, the question of complete reducibility is not
as straightforward. The matrix element formulae that will arise from these calculations will still be
valid, however, so what becomes challenging is to be able to {\em interpret} such
formulae in a particular circumstance. This problem will be addressed in a forthcoming
article. The current article, however, develops the important reduced matrix element formulae that
enable such full matrix element formulae to be expressed. In a sense, the work done in the current
article is analogous to our previous article \cite{GIW1}, which dealt with $gl(m|n)$.

There has been a substantial amount of interest in the representation theory of $osp(m|2k)$. Apart from
Jarvis and Green's \cite{JarGre1979,GreJar1983} and Gould's \cite{Gould1987} earlier work on characteristic identities for Lie
superalgebras, there are important works that deal with representations of orthosymplectic
superalgebras via their connection with parastatistics such as
\cite{Stoi2013,Palev1982,Van1984,LievStoiVan2008} and a wide variety of perspectives on developing
and understanding the representation theory
\cite{FarmJar1983,FarmJar1984,Nish1990,Quesne1990,BenLSRam1998,Coul2013,LS1999,LS2000,LS2002}.
Notably, the recent paper of \cite{Stoi2013} has exhibited a very constructive approach to the
representation theory of orthosymplectic Lie superalgebras, which is ultimately where our interest
lies. The current paper is therefore dedicated to laying the foundations for further work in
developing matrix element formulae similar in nature to the Gelfand-Tsetlin type formulae
\cite{GT1950,GT1950b}.

It is worth remarking that we have chosen to work with the canonical subalgebra chain
(\ref{subchain}) above, since it most naturally generalises the subalgebra chain used in the
representation theory of the orthogonal Lie algebras. The alternative of proceeding down the
subalgebra chain 
\begin{equation}
osp(m|2k+2)\supset osp(m|2k) \supset\cdots
\label{subchain2}
\end{equation}
gives rise to a state labelling problem
which will not be considered here.
As one can tell, however, continuing down the chain beyond what is written in
(\ref{subchain}) will take us to the symplectic Lie algebra $sp(2k)$. The problem for the
symplectic Lie algebra has been solved in a sense by Molev \cite{Molev1999} and also looked at by Gould
and Kalnins \cite{GouKal1985} and Gould \cite{Gould1989}. Note, however, that the subalgebra chain
(\ref{subchain2}) is particularly convenient when dealing with the case $osp(m|n)$ for $n\leq 4$, where no
state labelling problem arises.

The paper is organised as follows. After establishing notation, conventions, and some algebraic
structural aspects -- including root systems -- in Sections \ref{sec2} to \ref{sec8}, we look at
the representation theory of $osp(m|2k)$. In particular, in Section \ref{sec9} we introduce the
characteristic identities and in Section \ref{branchingcondition} give {\em branching conditions} as
a set of necessary conditions satisfied by the representations of the canonical subalgebra. 
After discussing vector operators in Section \ref{sec11},
we derive formulae for the reduced matrix elements in Sections \ref{sec12} and \ref{sec13}.


\section{$osp(m|n=2k)$ as a subalgebra of $gl(m|n)$}
\label{sec2}

Using the same fundamental notation as in \cite{GIW1,GIW2}, we have even indices $\{ i\ |\ 1\leq
i\leq m\}$, odd indices $\{\mu\ |\ 1\leq \mu\leq n=2k\}$, 
and the grading defined by
$$
(i)=0,\ \ (\mu)=1.
$$
Recall that $gl(m|n)$ is spanned by the elementary generators (in ungraded index notation)
$$
e_{pq},\ \ 1\leq p,q\leq m+n.
$$
We also recall, given an $(m+n)\times(m+n)$ homogeneous matrix $X$ so that
$$
X = 
\begin{blockarray}{cccc}
&&&\\
&&&\\
\begin{block}{(cc|cc)}
 &A & 0&\\ 
\cline{2-3}  
&0 & D&\\
\end{block}
&\underbrace{}_m & \underbrace{}_n& 
\end{blockarray}
,\ \ (X)=0,\ \ \mbox{ or }\ \ 
X = 
\begin{blockarray}{cccc}
&&&\\
&&&\\
\begin{block}{(cc|cc)}
 &0 & B&\\ 
\cline{2-3}  
&C & 0&\\
\end{block}
&\underbrace{}_m & \underbrace{}_n& 
\end{blockarray}
,\ \ (X)=1,
$$
the definition of {\em supertranspose}:
$$
\left( X^T\right)_{pq} = (-1)^{(X)(q)}X_{qp}.
$$
We note the following property:
$$
(XY)^T = (-1)^{(X)(Y)} Y^T X^T.
$$
To introduce the subalgebra $osp(m|n)$ we first need an even metric
\begin{equation}
g = \left( \begin{array}{c|c} g_0 & 0\\ \hline  0 & g_1 \end{array} \right)
	\label{gdef}
\end{equation}
with $g_0$ symmetric and $g_1$ anti-symmetric. Thus we have the symmetry
\begin{align}
	g_{pq} = (-1)^{(p)}g_{qp}
\label{gsym}
\end{align}
and note that $g_{pq}=0$ unless $(p)=(q)$. Letting $t$ denote the usual matrix transpose, 
it is also worth noting that
\begin{align*}
g^T &= g^t \\
&= \left( \begin{array}{c|c} g_0 & 0\\ \hline  0 & -g_1 \end{array} \right) = \gamma g = g
\gamma
\end{align*}
where 
\begin{equation}
\gamma = \left( \begin{array}{c|c} I & 0\\ \hline  0 & -I \end{array}
\right)
\label{gradaut}
\end{equation} 
is the grading automorphism.

\noindent
{\em Remark}: In general if $X$ is an even matrix, $X^T = X^t$, i.e. graded transpose
reduces to normal transpose for even operators.
$\blacksquare$ \medskip

Then it is easily seen that the set of matrices leaving the form $g$ invariant, i.e.
$$
L = \{ x\ |\ x^Tg + gx=0\ \}
$$
where $T$ is supertransposition, forms a Lie superalgebra, called the orthosymplectic Lie
superalgebra, denoted $osp(m|n)$.

To obtain a convenient basis for this Lie superalgebra we note that
\begin{align*}
x\in L \ \Leftrightarrow \ 0&= x^tg+gx\\
&= x^Tg^T\gamma + gx\\
&= (gx)^T\gamma + gx
\end{align*}
Then $x\in L$ $\Leftrightarrow$ $(gx)^T\gamma = -gx$ or, using $\gamma^2=I$,
$$
x\in L \ \Leftrightarrow \ (gx)^T = -gx\gamma.
$$
We call a homogeneous matrix $X$ anti-supersymmetric if
$$
X^T = -X\gamma
$$
with $\gamma$ the grading automorphism. Then we must have
$$
X_{pq} = -X_{qp}
$$
unless {\em both} indices are odd in which case $X_{pq}=X_{qp}$. Therefore $X$ is
anti-supersymmetric if and only if
$$
X_{pq} = -(-1)^{(p)(q)}X_{qp}.
$$

The anti-supersymmetric matrices have a basis consisting of the matrices
$$
e_{pq}-(-1)^{(p)(q)}e_{qp},\ \ 1\leq p,q\leq m+n,
$$
i.e.
\begin{align*}
e_{ij}-e_{ji},&\ \ 1\leq i<j\leq m,\\
e_{i\mu} - e_{\mu i},&\ \ 1\leq i\leq m, \ 1\leq \mu\leq n,\\
e_{\mu\nu} + e_{\nu\mu},&\ \ 1\leq \mu,\nu\leq n.
\end{align*}
This gives rise to the $osp(m|n)$ generators:
\begin{align*}
  x\in L=osp(m|n) &\Leftrightarrow \ gx \mbox{ is anti-supersymmetric}\\
  &\Leftrightarrow \ x = g^{-1}\left( e_{pq}-(-1)^{(p)(q)}e_{qp} \right).
\end{align*}
Under the assumption $g$ is orthogonal, so $g^t=g^{-1}$, which we will do, we obtain the
$osp(m|n)$ generators
\begin{align}
	\sigma_{pq} &= g^t\left( e_{pq}-(-1)^{(p)(q)}e_{qp} \right) \nonumber\\
				  &= g^t_{rp}e_{rq} - (-1)^{(p)(q)}g^t_{rq}e_{rp} \nonumber \\
      &= g_{pr}e_{rq} - (-1)^{(p)(q)}g_{qr}e_{rp}, \label{ospgen}
\end{align}
where we have adopted the summation convention over repeated indices.
Here $e_{pq}$ are the $gl(m|n)$ generators satisfying the graded commutation relations
$$
[e_{pq},e_{rs}] = \delta_{rq}e_{ps} - (-1)^{((p)+(q))((r)+(s))}\delta_{ps}e_{rq}.
$$
Using this it can be shown that the generators (\ref{ospgen}) satisfy the symmetry
property
$$
\sigma_{pq} = -(-1)^{(p)(q)}\sigma_{qp}
$$
as well as the graded commutation relations
\begin{align}
	[\sigma_{pq},\sigma_{rs}] &= g_{rq}\sigma_{ps}-(-1)^{((p)+(q))((r)+(s))}g_{ps}\sigma_{rq}
\nonumber \\
& \ \ -(-1)^{(p)(q)}\left\{ g_{rp}\sigma_{qs}-(-1)^{((p)+(q))((r)+(s))}g_{qs}\sigma_{rp}
\right\}.
\label{osprels}
\end{align}

We shall also work with the generators
$$
\sigma^p_{\ q} = g^{pr}\sigma_{rq}
$$
where $g^{pr}\equiv \left(g^{-1}\right)_{pr}$. This gives the generators
\begin{align}
\sigma^p_{\ q} &= g^{pr}\sigma_{rq} \nonumber \\
&= g^{pr}\left\{ g_{rs}e_{sq} - (-1)^{(r)(q)}g_{qs}e_{sr} \right\} \nonumber\\
&= e_{pq} - (-1)^{(p)(q)}g^{pr}g_{qs}e_{sr} \label{altospgen}
\end{align}
which will depend on the explicit choice of metric $g$.

%
%

\section{Adjoint tensors} \label{sec3}

Operators $X_{pq}$ are said to form an $osp(m|n)$ adjoint tensor operator if they satisfy
the same relations as the $\sigma_{pq}$ under graded commutation:
\begin{align}
	[\sigma_{pq},X_{rs}] &= g_{rq}X_{ps}-(-1)^{((p)+(q))((r)+(s))}g_{ps}X_{rq}
\\
& \ \ -(-1)^{(p)(q)}\left\{ g_{rp}X_{qs}-(-1)^{((p)+(q))((r)+(s))}g_{qs}X_{rp}
\right\}
\label{adjtensor}
\end{align}
where the left hand side is the usual graded commutator. We immediately have the
following.

\begin{lemma}
Let $X_{pq}$ be an adjoint tensor. Then
\begin{itemize}
\item[(i)] The trace
$$
x \equiv X^p_{\ p} =\mbox{tr}(X) = g^{pq}X_{qp}
$$
is always an invariant, i.e. $[\sigma_{pq},x]=0$.
\item[(ii)] If $W_{pq}$ is another adjoint tensor, then so too is
$$
V_{pq}\equiv X_{pr}(-1)^{(r)}g^{rs}W_{sq}.
$$
\end{itemize}
\end{lemma}
\proof{
The result of (i) is easily obtained by substituting $x = g^{pq}X_{qp}$ into the
graded commutator $[\sigma_{pq},x]$ and making use of the symmetry property (\ref{gsym})
of the metric, along with $g^{sr}g_{rq} = \delta^s_{\ r}$. The result of (ii) also follows
by straightforward application of these properties, in addition to the graded derivation rule
$$
[A,BC] = [A,B]C + (-1)^{(A)(B)}B[A,C].
$$
}

The above suggests we define matrix powers
\begin{align*}
	\hat{\sigma}^p_{\ q} &= (-1)^{(p)}g^{pr}\sigma_{rq} = (-1)^{(p)}\sigma^p_{\ q},\\
\left( \hat{\sigma}^{N+1} \right)^p_{\ q} &= \hat{\sigma}^p_{\ r}\left( \hat{\sigma}^N \right)^r_{\ q}
= \left( \hat{\sigma}^N \right)^p_{\ r} \hat{\sigma}^r_{\ q}
\end{align*}
with $\left( \hat{\sigma}^0 \right)^p_{\ q} = \delta^p_{\ q}.$ Note that as a matrix
$\hat{\sigma}=\gamma\sigma,$ with $\gamma$ the grading automorphism. Then it is easily
seen that 
$$
X_{pq} = (-1)^{(p)}g_{pr}\left( \hat{\sigma}^N \right)^r_{\ q}
$$
transforms as an adjoint tensor, so we arrive at the Casimir invariants
\begin{align*}
I_N &=  g^{pq}X_{qp}\\
    &=  g^{pq}(-1)^{(q)} g_{qr}\left( \hat{\sigma}^N \right)^r_{\ p}\\
    &=  (-1)^{(p)} \left( \hat{\sigma}^N \right)^p_{\ p}\\
&= \mbox{str}\left( \hat{\sigma}^N \right). 
\end{align*}
In particular we have the second order invariant
$$
I_2 = \mbox{str}\left( \hat{\sigma}^2 \right)
$$
which we will see is actually twice the universal Casimir element, while the first order
invariant vanishes identically. Indeed
\begin{align*}
I_1 &= \mbox{str}\left( \hat{\sigma} \right)\\
 &= g^{pq}\sigma_{qp} \\
 &= g^{pq}\left( g_{qr}e_{rp} - (-1)^{(p)(q)}g_{pr}e_{rq} \right)\\
&= \left( e^p_{\ p} - g^{qp}g_{pr}e_{rq} \right)\\
&= e^p_{\ p} - e^q_{\ q} = 0.
\end{align*}


\section{Racah generators} \label{sec4}

To obtain the relations satisfied by the $osp(m|n)$ generators $\sigma^p_{\ q}$ we need to
make a choice for our metric $g$. We choose $g$ to give the Racah $o(m)$ generators which
respect the subalgebra embedding
$$
osp(m|n)\supset osp(m-1|n).
$$
Thus we take $g$ as in (\ref{gdef}) with
$$
g_0 = I \mbox{ (i.e. $m\times m$ identity matrix),}
$$
and
$$
g_1 =
\begin{blockarray}{cc}
\begin{block}{(c)c}
  \begin{array}{c|c} 
    0 & 
    \begin{array}{cccc} &&&1\\ && 1 & \\ & \iddots & & \\ 1 & && \end{array}
    \\ 
    \cline{1-2} 
    \begin{array}{cccc} &&&-1\\ && -1 & \\ & \iddots & & \\ -1 & && \end{array}
    & 0  
  \end{array}
&
  \begin{array}{c} 
    \left.\vphantom{\begin{array}{cccc} &&&1\\ && 1 & \\ & \iddots & & \\ 1 & &&
    \end{array}}\right\} k
    \\ 
    \vphantom{\begin{array}{cccc} &&&-1\\ && -1 & \\ & \iddots & & \\ -1 & && \end{array}}
  \end{array}
\\
\end{block}
\underbrace{
\hphantom{ 
\begin{array}{cc|l} 
0 & 
\left. \begin{array}{cccc} &&&1\\ && 1 & \\ & \iddots & & \\ 1 & && \end{array}
\right\} & k \\ 
\cline{1-2} 
\begin{array}{rrrr} &&&-1\\ && -1 & \\ & \iddots & & \\ -1 & && \end{array}
& 0 & \end{array}
} 
}_n & 
\end{blockarray}.
$$
Hence with this choice we have
\begin{align*}
g_{ij} &= \delta_{ij},\ \ 1\leq i,j\leq m,\\
\delta_{\mu\nu} &= \theta_\mu \delta_{\mu\overline{\nu}},\ \ \overline{\nu}-n+1-\nu,\
1\leq \mu,\nu\leq n,\\
g_{i\mu} &= g_{\mu i} = 0,\ \ 1\leq i\leq m,\ 1\leq \mu\leq n,
\end{align*}
where $\theta_{\mu}$ is the step function defined by
$$
\theta_\mu = 
\left\{
\begin{array}{rl} 1, & \mu \leq k=n/2,\\
-1, & \mu > k.
\end{array}
\right.
$$
Then $g$ is indeed orthogonal since
$$
g^{-1} = \left( \begin{array}{c|c} I & 0\\ \hline 0 & g_1^t \end{array} \right) = g^t
$$
and satisfies 
\begin{align*}
g^2 &= \left( \begin{array}{c|c} I & 0\\ \hline 0 & -I \end{array} \right) = \gamma \\
\Rightarrow g^{-1} &= \gamma g = g\gamma.
\end{align*}
Thus our generators $\sigma^p_{\ q}$ are given, from (\ref{altospgen}), by
$$
\sigma^p_{\ q} = e_{pq} - (-1)^{(p)(q)}g^{pr}g_{qs}e_{sr}.
$$
Now observe that 
\begin{align*}
g^{pq} &= \left( g^{-1} \right)_{pq} = \left( g^t \right)_{pq} \\
&= g_{qp} = \theta_q\delta_{p\tilde{q}}
\end{align*}
where we now define for even indices $p=i$, $\theta_i=1$ and the opposite index
$\tilde{p}$ is given by
$$
\tilde{p} = \left\{
\begin{array}{rl}
i, &p=i,\\
\overline{\mu}, & p = \mu,
\end{array}
\right.
$$
i.e. $\tilde{i} = i,$ $\tilde{\mu} = \overline{\mu}$. In this notation we have
\begin{align*}
	\sigma^p_{\ q} &= e_{pq} - (-1)^{(p)(q)}g^{pr}g_{qs} e_{sr} \\
	&= e_{pq} - (-1)^{(p)(q)}\theta_r\delta_{p\tilde{r}}\theta_q\delta_{q\tilde{s}} e_{sr}\\
 &= e_{pq} - (-1)^{(p)(q)}\theta_{\tilde{p}}\theta_q e_{\tilde{q}\tilde{p}}.
\end{align*}
Finally observe that 
$$
\theta_p\theta_{\tilde{p}} = (-1)^{(p)},\ \ \theta_p^2=1,
$$
so we arrive at
\begin{align}
	\sigma^p_{\ q} = e_{pq} - (-1)^{(p)((p)+(q))}\theta_p\theta_q e_{\tilde{q}\tilde{p}}
\label{Racahgen}
\end{align}
which are the Racah generators. They satisfy the symmetry relation
\begin{align}
	\sigma^p_{\ q} = -(-1)^{(p)((p)+(q))} \theta_p\theta_q \sigma^{\tilde{q}}_{\ \tilde{p}}.
\label{Racahsymm}
\end{align}
{\em Remark}:  Explicitly we have
\begin{align*}
\sigma^i_{\ j} &= e_{ij} - e_{ji}, \mbox{ (Racah generators of $o(m)$) } \\
\sigma^{\mu}_{\ \nu} &= e_{\mu\nu} -
\theta_{\mu}\theta_{\nu}e_{\overline{\nu}\overline{\mu}} \\
\sigma^i_{\mu} &= e_{i\mu} - \theta_\mu e_{\overline{\mu} i}\\
\sigma^{\mu}_i &= e_{\mu i} + \theta_\mu e_{i\overline{\mu}}.
\end{align*}
$\blacksquare$ \medskip

The above Racah generators satisfy the following graded commutation relations
\begin{align}
[\sigma^p_{\ q},\sigma^r_{\ s}] &= \delta^r_{\ q}\sigma^p_{\ s} -
	(-1)^{((p)+(q))((r)+(s))} \delta^p_{\ s}\sigma^r_{\ q}\nonumber\\
	&\ \ - (-1)^{(p)((p)+(q))}\theta_p\theta_q\left\{ \delta^r_{\
\tilde{p}}\sigma^{\tilde{q}}_{\ s} - (-1)^{((p)+(q))((r)+(s))}\delta^{\tilde{q}}_{\
s}\sigma^r_{\ \tilde{p}} \right\}
\label{Racahrels}
\end{align}
as we would expect in view of the symmetry relation (\ref{Racahsymm}).


\section{Cartan-Weyl generators} \label{sec5}

The trouble with the above Racah generators is that they are not in Cartan-Weyl form.
Thus it is necessary to find a Cartan subalgebra for $osp(m|n)$ in the above basis and
furthermore to find the Cartan-Weyl generators in terms of the Racah generators. This
problem was first considered for the Lie algebra $o(m)$ by Wong \cite{Wong1967} and Pang and Hecht
\cite{PangHecht1967}. 

Here we consider a simplified approach to this problem based on previous work of Gould
\cite{Gould1978}. Following the latter we introduce the numerical $o(m)$ matrix which, for
$m=2h$ even, is given by
$$
M = \frac{1}{\sqrt{2}}
\begin{blockarray}{rrrrrcrrrr}
&&&&&&&&&\\
\begin{block}{(rrrrrcrrrr)}
1       &  0     & 0        & \cdots &  0     & 0      & \cdots & 0       & 0      & 1       \\
-i      &  0     & 0        &        & \vdots & \vdots &        & 0       & 0      & i       \\
 0      &  1     & 0        &        &        &        &        & 0       & 1      & 0       \\
 0      & -i     & 0        & \cdots &        &        & \cdots & 0       & i      & 0       \\
 \vdots &        & \ddots   &        &        &        &        & \iddots &        &  \vdots \\
        &        &          &   1    & 0      & 0      &   1    &         &        &         \\
        &        &          &  -i    & 0      & 0      &   i    &         &        &         \\
 \vdots &        &          &   0    & 1      & 1      &   0    &         &        &  \vdots \\
   0    & \cdots &          &   0    & -i     & i      &   0    &         & \cdots &     0   \\
\end{block}
       &&&&        \stackrel{\uparrow}{{\tiny \mbox{$h$} }} & \stackrel{\uparrow}{{\tiny
\mbox{$h+1$} }}    &&&& 
\end{blockarray}
$$
i.e. for $1\leq j\leq h$, 
\begin{align*}
M^{2j-1}_{\phantom{2j-1}\ j} &= 1/\sqrt{2} = M^{2j-1}_{\ \ m+1-j},\\
M^{2j}_{\phantom{2j}\ j} &= -i/\sqrt{2} = -M^{2j}_{\ \ m+1-j},
\end{align*}
with all other entries being zero. When $m=2h+1$ is odd we add an extra row and column to
give
$$
M = \frac{1}{\sqrt{2}}
\begin{blockarray}{rrrrrcrrrrr}
&&&&&&&&&&\\
\begin{block}{(rrrrrcrrrrr)}
1       &  0     & 0        & \cdots & 0  & 0      & 0 & \cdots & 0       & 0      & 1       \\
-i      &  0     & 0        &        &    & \vdots   &   &        & 0       & 0      & i       \\
 0      &  1     & 0        &        &    &        &   &        & 0       & 1      & 0       \\
 0      & -i     & 0        & \cdots &    &        &   & \cdots & 0       & i      & 0       \\
 \vdots &        & \ddots   &        &    &        &   &        & \iddots &        &  \vdots \\
        &        &          &   1    & 0  &        & 0 &   1    &         &        &         \\
        &        &          &  -i    & 0  &        & 0 &   i    &         &        &         \\
 \vdots &        &          &   0    & 1  & \vdots & 1 &   0    &         &        &  \vdots \\
   0    & \cdots &          &   0    & -i &    0   & i &   0    &         & \cdots &     0   \\
   0    & \cdots &          &       & 0 &    \sqrt{2}     & 0 &        &         & \cdots &     0   \\
\end{block}
       &&&&        & \stackrel{\uparrow}{{\tiny \mbox{$h+1$} }}  &  &&&& 
\end{blockarray}
$$
so we have an additional non-zero entry $\displaystyle{M^m_{\ h+1}=1.}$

We note that the columns of the matrix $M$ form an orthonormal basis so that $M$ is a
unitary matrix, i.e.
$$
M^\dagger =  M^{-1}
$$
as may be easily shown. We note also the following additional symmetry:
\begin{align}
\overline{M}_{ij} = M_{i\overline{j}},\ \ 1\leq i,j\leq m
\label{Msymm}
\end{align}
where now $\overline{i}$ (the opposite index) is given by $\overline{i} = m+1-i.$
Combining these we have:

\begin{lemma} \label{lemmap11}
$$
M_{k\overline{i}} M_{kj} = \delta_{ij}
$$
i.e. the columns of $M$ are orthogonal in this sense. Similarly
$$
\left( M^{-1} \right)_{\overline{i}k}\left( M^{-1} \right)_{jk} = \delta_{ij}.
$$
\end{lemma}
\proof{
Since $M^{-1}=M^\dagger$ we have
\begin{align*}
\delta_{ij} &= \left(M^\dagger\right)_{ik} M_{kj} = \overline{M}_{ki}M_{kj}\\
&\stackrel{(\ref{Msymm})}{=} M_{k\overline{i}}M_{kj},
\end{align*}
as required. Similarly for the second statement.
}

We extend $M$ to an $(m+n)\times(m+n)$ matrix in a trivial way by defining
$$
M_{\mu\nu} = \delta_{\mu\nu},\ \ M_{i\mu} = M_{\mu i} = 0.
$$
We now introduce the Cartan-Weyl generators
\begin{align*}
S^p_{\ q} &= \left(M^{-1} \sigma M\right)^p_{\ q}\\
&= \left(M^{-1}\right)^p_{\ r}\sigma^r_{\ s} M^s_{\ q}
\end{align*}
where $M^{-1}$ satisfies
$$
\left( M^{-1} \right)_{i\mu} = \left( M^{-1}\right)_{\mu i} = 0,\ \ \left(
M^{-1}\right)_{\mu\nu} = \delta_{\mu\nu}.
$$
In view of equation (\ref{Racahrels}) we have the graded commutation relations
\begin{align*}
[S^p_{\ q},S^r_{\ s}] &= \left( M^{-1} \right)^p_{\ p'}\left( M^{-1} \right)^r_{\
r'}[\sigma^{p'}_{\ q'},\sigma^{r'}_{\ s'}]M^{q'}_{\ q}M^{s'}_{\ s} \\
&=\left( M^{-1} \right)^p_{\ p'}\left( M^{-1} \right)^r_{\ r'}  
\left[ \delta^{r'}_{\ q'}\sigma^{p'}_{\ s'} - (-1)^{((p)+(q))((r)+(s))} \delta^{p'}_{\
s'}\sigma^{r'}_{\ q'} \right. \\
&\ \ \left. -(-1)^{(p)((p)+(q))}\theta_p\theta_q\left\{ \delta^{r'}_{\ \tilde{p}'}
\sigma^{\tilde{q}'}_{\ s'} - (-1)^{((p)+(q))((r)+(s))}\delta^{\tilde{q}'}_{\
s'}\sigma^{r'}_{\ \tilde{p}'} \right\}  \right] M^{q'}_{\ q}M^{s'}_{\ s} \\
&= \delta^r_{\ q}S^p_{\ s} - (-1)^{((p)+(q))((r)+(s))}\delta^p_{\ s}S^r_{\ q}\\
&\ \ -(-1)^{(p)((p)+(q))} \theta_p\theta_q\left\{  \left( M^{-1} \right)^p_{\ p'}
\left( M^{-1} \right)^r_{\ \tilde{p}'}\sigma^{\tilde{q}'}_{\ s'}M^{q'}_{\ q}M^{s'}_{\ s}
\right. \\
& \ \ \ \ \left.-(-1)^{((p)+(q))((r)+(s))}\left( M^{-1} \right)^p_{\ p'}\left( M^{-1} \right)^r_{\
r'} \sigma^{r'}_{\ \tilde{p}'} M^{q'}_{\ q}M^{\tilde{q}'}_{\ s} \right\}.
\end{align*}
Using Lemma \ref{lemmap11} we have
$$
\left(M^{-1}\right)^p_{\ p'}  \left( M^{-1} \right)^r_{\ \tilde{p}} = 
\left\{ 
	\begin{array}{rl} 
		\delta_{r\overline{p}}, & p,r \mbox{ even,}\\
		\delta_{r\overline{p}}, & p,r \mbox{ odd.}
	\end{array}
\right.  
$$
Thus in either case,
\begin{align}
	\left(M^{-1}\right)^p_{\ p'}  \left( M^{-1} \right)^r_{\ \tilde{p}} =
	\delta_{r\overline{p}}. \label{starp13}
\end{align}
Similarly using
$$
S = M^{-1}\sigma M\ \ \Rightarrow\ \ \sigma M = MS
$$
we have 
\begin{align*}
\sigma^{\tilde{q}'}_{\ s'} M^{q'}_{\ q}M^{s'}_{\ s} 
&= (\sigma M)^{\tilde{q}'}_{\ s}M^{q'}_{\ q}\\
&=  (MS)^{\tilde{q}'}_{\ s}M^{q'}_{\ q}\\
&=  M^{\tilde{q}'}_{\ s'}S^{s'}_{\ s}M^{q'}_{\ q}
\end{align*}
where now
$$
M^{\tilde{q}'}_{\ s'}M^{q'}_{\ q} = \delta_{\overline{q}s'}
$$
(c.f. equation (\ref{starp13}) above).
In a similar way we may show that
$$
\left( M^{-1} \right)^p_{\ p'}  \left( M^{-1} \right)^r_{\ r'}\sigma^{r'}_{\ \tilde{p}'}
=
\left( M^{-1}\right)^p_{\ p'} S^r_{\ r'} \left( M^{-1} \right)^{r'}_{\ \tilde{p}'}.
$$
{\em Remark}: For index $p$ we now define the opposite index $\overline{p}$ by
$$
\overline{p} = \left\{ 
	\begin{array}{rl}
		\overline{i} = m+1-i,& p=i \mbox{ even,}\\
		\overline{\mu} = n+1-\mu,& p=\mu \mbox{ odd.}
	\end{array}
\right.
$$
$\blacksquare$ \medskip

Substituting into the previous relation we arrive at the graded commutation relations
\begin{align}
[S^p_{\ q},S^r_{\ s}] &=  
	\delta^r_{\ q}S^p_{\ s} - (-1)^{((p)+(q))((r)+(s))}\delta^p_{\ s}S^r_{\ q}\\
	&\ \ -(-1)^{(p)((p)+(q))} \theta_p\theta_q\left\{ \delta^r_{\
\overline{p}}S^{\overline{q}}_{\ s} - (-1)^{((p)+(q))((r)+(s))}\delta^{\overline{q}}_{\
s}S^r_{\ \overline{p}}  \right\}.
\label{secondstarp13}
\end{align}

Recall that $\overline{i}=m+1-i,$ $\overline{\mu}=n+1-\mu,$ $\theta_i=1,$
$$
\theta_\mu \ \left\{ \begin{array}{rl} 1,& \mu\leq k,\\ -1,& \mu>k \end{array}\right.
$$
and $\theta_p^2=1,$ $\theta_p\theta_{\overline{p}} = (-1)^{(p)}$.
By case splitting it is easily seen that the generators $S^p_{\ q}$ satisfy the symmetry
relation
$$
S^p_{\ q} = -(-1)^{(p)((p)+(q))} \theta_p \theta_q S^{\overline{q}}_{\ \overline{p}}.
$$
In particular, setting $p=q$ into equation (\ref{secondstarp13}) we obtain
\begin{align}
	[S^p_{\ p},S^r_{\ s}] = \left( \delta^r_{\ p} - \delta^p_{\ s} - \delta^r_{\
	\overline{p}} + \delta^{\overline{p}}_{\ s} \right) S^r_{\ s} \label{starp14}
\end{align}
$$
\Rightarrow \ \ [S^p_{\ p},S^r_{\ r}]=0,\ \ S^p_{\ p} = -S^{\overline{p}}_{\
\overline{p}}.
$$
Thus the diagonal generators commute and thus span a Cartan subalgebra, while relations
(\ref{starp14}) show that our generators are indeed in Cartan-Weyl form. 

To be explicit, as a Cartan subalgebra, we take the diagonal generators
\begin{align*}
	S^i_{\ i} &= -S^{\overline{i}}_{\ \overline{i}},\ \ 1\leq i\leq h = \left\lfloor
	\frac{m}{2}\right\rfloor\\
	S^\mu_{\ \mu} &= -S^{\overline{\mu}}_{\ \overline{\mu}},\ \ 1\leq \mu\leq k.
\end{align*}
{\em Notes:}
\begin{itemize}
	\item[(1)] Thus for $m=2h+1$ (an odd number) we have for the index $i=h+1$,
		$\overline{h+1}=h+1$
		$$
		\Rightarrow \ \ S^{h+1}_{\ h+1} = -S^{h+1}_{\ h+1} = 0,
		$$
i.e., this generator is identically zero in that case.
\item[(2)]
	Further for even indices we always have
	$$
	S^i_{\ \overline{i}} = S^{\overline{i}}_{\ i} = 0,\ \ 1\leq i\leq h
	$$
	which is consistent with the fact that the weight $2\varepsilon_i$ is not a root
	for $o(m)$.
\end{itemize}


\section{Root system, positive roots and $\mathbb{Z}$-grading ($m\geq 2$)} \label{sec6}

In ungraded index notation it follows that the generator $S^r_{\ s}$ has weight
$\varepsilon_r-\varepsilon_s$, where we define weights $\varepsilon_p$ by
$$
\varepsilon_{\overline{i}} = -\varepsilon_i,\ \ 1\leq i\leq h,
$$
with $\varepsilon_{h+1}=0$ for $m=2h+1$ and
$$
\varepsilon_\mu = \delta_\mu,\ \ \delta_{\overline{\mu}} = -\delta_\mu,\ \ 1\leq \mu\leq
k.
$$
Above, $\varepsilon_i$ ($1\leq i\leq h$), $\delta_\mu$ ($1\leq\mu\leq k$) denote our
elementary even and odd weights with 1 in position $i$ (respectively $\mu$) and zeros
elsewhere.

Thus our roots are $\varepsilon_r-\varepsilon_s\neq0$ or, in graded index notation, we
have the {\em even} roots
\begin{align*}
&\pm\varepsilon_i\pm\varepsilon_j,\ \ 1\leq i<j\leq h,\\
&\pm2\delta_\mu,\ \ \pm\delta_\mu\pm\delta_\nu,\ \ 1\leq\mu<\nu\leq k,
\end{align*}
together with $\pm\varepsilon_i$ ($1\leq i\leq h$) for odd $m=2h+1$. Our {\em odd} roots
are
$$
\pm\varepsilon_i\pm\delta_\mu,\ \ 1\leq i\leq h,\ 1\leq \mu\leq k
$$
together with
$$
\pm\delta_\mu = \varepsilon_{h+1}\pm\delta_\mu
$$
when $m=2h+1$ is odd.

To obtain the corresponding {\em positive} roots we need to look at the
$\mathbb{Z}$-graded structure of $L=osp(m|n)$. 
Here our {\em odd} positive roots are given by the weights
$$
\varepsilon_i+\delta_\mu,\ \ 1\leq i\leq m,\ 1\leq\mu\leq k.
$$
In the case $m=2h+1$ is odd, this includes the odd roots
$\delta_\mu=\varepsilon_{h+1}+\delta_\mu,$ $1\leq\mu\leq k$. Thus the set of odd positive
roots is given by 
$$
\Phi_1^+ = \{\varepsilon_i+\delta_\mu\ |\ 1\leq i\leq m,\ 1\leq \mu\leq k\}.
$$
The even positive roots are given by $\Phi_{\overline{0}}^+$ which is the set of (even)
positive roots of the even subalgebra $L_{\overline{0}} = o(m)\oplus sp(n)$. Thus our even
positive roots are given by
\begin{align*}
&\delta_\mu\pm\delta_\nu\ \ (1\leq\mu<\nu\leq k),\ \ 2\delta_\mu\ \ (1\leq\mu\leq k)\\
&\varepsilon_i\pm\varepsilon_j,\ \ 1\leq i<j\leq h,
\end{align*} 
together with $\varepsilon_i$, $1\leq i\leq h$ in the case $m=2h+1$ is odd. 

We first observe that $L$ admits a $\mathbb{Z}_2$-grading
$L=L_{\overline{0}}\oplus L_{\overline{1}}$ where $L_{\overline{0}}=o(m)\oplus sp(n)$ is
the underlying even sub-Lie algebra and 
\begin{align*}
L_{\overline{1}} &= \mbox{span}\left\{ S^i_{\ \mu}\ |\ 1\leq i\leq m,\ 1\leq\mu\leq n
\right\}\\
&= \mbox{span}\left\{ S^\mu_{\ i}\ |\ 1\leq i\leq m,\ 1\leq \mu\leq n \right\}.
\end{align*}  
We moreover have a $\mathbb{Z}$-grading
\begin{align}
L = L_{-2}\oplus L_{-1}\oplus L_0\oplus L_1\oplus L_2
\label{starp17}
\end{align}
where
\begin{align*}
L_2=\mbox{span}\left\{ S^\mu_{\ \overline{\nu}}\ |\ 1\leq\mu,\nu\leq k \right\},
L_{-2}=\mbox{span}\left\{ S^{\overline{\mu}}_{\ {\nu}}\ |\ 1\leq\mu,\nu\leq k \right\}.
\end{align*}
Also,
$$
L_{\overline{0}} = L_{-2}\oplus L_0\oplus L_2
$$
where $L_0 = o(m)\oplus gl(k)$ and
$$
gl(k) = \mbox{span}\left\{ S^{\mu}_{\ \nu}\ |\ 1\leq\mu,\nu\leq k \right\}.
$$
Then 
$$
L_{\overline{1}} = L_{-1}\oplus L_1
$$
with 
\begin{align*}
L_1 &= \mbox{span}\left\{ S^\mu_{\ i}\ |\ 1\leq i\leq m,\ 1\leq \mu\leq k \right\}\\
L_{-1} &= \mbox{span}\left\{ S^i_{\ \mu}\ |\ 1\leq i\leq m,\ 1\leq \mu\leq k \right\}.
\end{align*}

Note that the set of even positive roots is given by
$$
\Phi_{\overline{0}}^+ = \Phi_0^+ \cup \left\{ \delta_\mu+\delta_\nu\ | \
1\leq\mu\leq\nu\leq k \right\}
$$
with $\Phi_0^+$ the (positive) roots of $L_0 = o(m)\oplus gl(k)$.

The result of this discussion is that the above choice of positive roots is consistent with the
$\mathbb{Z}$-grading (\ref{starp17}) on $L=osp(m|n)$. 

\noindent
{\em Note}: 
In the case $m=2$, corresponding to the case $L=C(k+1)$, the above $\mathbb{Z}$-grading and
hence the choice of simple roots -- particularly the odd positive roots -- is quite different to Kac's distinguished choice. See Appendix C for
details.


\section{Highest weights, irreducible $L$-modules and $\mathbb{Z}$-gradings} \label{sec8}

Consistent with the $\mathbb{Z}$-gradation (\ref{starp17}), every finite dimensional irreducible
$L$-module $V(\Lambda)$, with highest weight $\Lambda$, admits a $\mathbb{Z}$-gradation 
$$
V(\Lambda) = \bigoplus_{i=-d}^0 V_i(\Lambda)
$$
whose maximal $\mathbb{Z}$-graded component $V_0(\Lambda)$ is necessarily a finite dimensional
irreducible $L_0$-module. It follows that $\Lambda$ must be dominant for $L_0=o(m)\oplus gl(k)$.
Furthermore, $\Lambda$ must be the highest weight of an irreducible $L_{\overline{0}}=o(m)\oplus
sp(n)$  module so in fact $\Lambda$ must be dominant for $L_{\overline{0}}$.

It follows that the components of 
$$
\Lambda = \sum_{i=1}^h\Lambda_i\varepsilon_i +\sum_{\mu=1}^k\Lambda_\mu\delta_\mu
$$
must satisfy
$$
\Lambda_{\mu=k}\geq\Lambda_{k-1}\geq \cdots\geq\Lambda_{\mu=1}\geq 0,\ \ \Lambda_\mu\in\mathbb{Z}
$$
(dominant condition for $sp(n)$) and 
\begin{align*}
&\Lambda_{i=1}\geq \Lambda_2\geq \cdots\geq \Lambda_{i=h}\geq 0,\ \ m=2h+1,\\
&\Lambda_1\geq \Lambda_2\geq \cdots\geq |\Lambda_h|,\ \ m=2h,\\
& 2\Lambda_i\in\mathbb{Z},\ \ \Lambda_i-\Lambda_j\in\mathbb{Z},\ \ i\neq j,
\end{align*}
so that the even components of $\Lambda$ are simultaneously either all integers (tensor
representations) or all $1/2$-odd integers (spinor representations). Thus, unlike type I Lie
superalgebras (in the sense of Kac), we do {\em not} in general obtain one-parameter families of irreducible modules.

The eigenvalues of the universal Casimir element $C_L$ on $V(\Lambda)$ are well known to be given by
\begin{align}
\chi_\Lambda(C_L) = (\Lambda,\Lambda+2\rho)
\label{p19star}
\end{align}
where $\rho = \rho_0-\rho_1$,
\begin{align}
\rho_0 = \frac12\sum_{\alpha\in\Phi^+_{\overline{0}}} \alpha,\ \ \rho_1 = \frac12
\sum_{\beta\in\Phi^+_1}\beta,
\label{rhos}
\end{align}
is the graded $1/2$-sum of the positive roots, and $(\ ,\ )$ is the bilinear form induced on our
weights defined by
\begin{align}
(\varepsilon_i,\varepsilon_j) = \delta_{ij},\ \ (\delta_\mu,\delta_\nu) = -\delta_{\mu\nu},\ \
(\varepsilon_i,\delta_\mu)=0,\ \ 1\leq i,j\leq h,\ 1\leq \mu,\nu\leq k.
\label{signs}
\end{align}
We have explicitly
\begin{align}
\rho_0 &= \frac12\sum_{i=1}^h(m-2i)\varepsilon_i + \frac12\sum_{\mu=1}^k(n-2\mu+2)\delta_\mu,
\nonumber\\
\rho_1 &= \frac{m}{2}\sum_{\mu=1}^k \delta_\mu 
\label{p19twostar}
\end{align}
\begin{align}
\Rightarrow \ \ \rho = \frac12\sum_{i=1}^h(m-2i)\varepsilon_i +
\frac12\sum_{\mu=1}^k(n-m-2\mu+2)\delta_\mu.
\label{starp20}
\end{align}

To see how the sign in the definition of the bilinear form $(\ ,\ )$ arises, it is instructive to
consider the quadratic invariant
\begin{align*}
I_2 &= \mbox{str}(\hat{\sigma}^2)\\
    &= \sigma^p_{\ q}(-1)^{(q)}\sigma^q_{\ p}\\
    &= S^p_{\ q}(-1)^{(q)}S^q_{\ p}\\
    &= (-1)^{(p)}\left({\left(\gamma S\right)^2}\right)^p_{ \ p} = \mbox{str}\left( \hat{S}^2 \right),
\end{align*}
where 
$$
\hat{S}=\gamma S \ \ \Rightarrow \ \ \hat{S}^p_{\ q} = (-1)^{(p)}S^p_{\ q}.
$$
Thus if $v^\Lambda_+$ is the maximal weight vector of $V(\Lambda)$, we have (summation over repeated
indices)
\begin{align*}
I_2v^\Lambda_+ &= \left\{ S^p_{\ i}S^i_{\ p} - S^p_{\ \mu}S^\mu_{\ p} \right\}v^\Lambda_+\\
&= S^i_{\ j}S^j_{\ i}v^\Lambda_+ - S^\mu_{\ \nu}S^\nu_{\ \mu}v^\Lambda_+ +S^\mu_{\ i}S^i_{\
\mu}v^\Lambda_+ - S^i_{\ \overline{\mu}} S^{\overline{\mu}}_{\ i}v^\Lambda_+,
\end{align*}
where we have used the fact that the raising generators vanish on $v^\Lambda_+$. Now 
$$
S^i_{\ j}S^j_{\ i} = 2C_{L_0}
$$
is twice the Casimir element of $o(m)$, so 
$$
S^i_{\ j}S^j_{\ i}v^\Lambda_+ = 2(\Lambda_0,\Lambda_0+2\rho_0)v^\Lambda_+
$$
where 
$$
\Lambda_0 = \sum_{i=1}^h\Lambda_i\varepsilon_i
$$
is the projection of $\Lambda$ onto the space of even weights. Similarly setting
$$
\Lambda_1 = \sum_{\mu}^k\Lambda_\mu\delta_\mu
$$
we have
$$
S^\mu_{\ \nu}S^\nu_{\ \mu}v^\Lambda_+ = 2(\Lambda_1,\Lambda_1+2\rho_1)v^\Lambda_+
$$
since $S^\mu_{\ \nu}S^\nu_{\ \mu}$ is twice the universal Casimir element of $sp(n)$ and where the
negative sign needs to be incorporated in the definition of the form $(\ ,\ )$ as in (\ref{signs})
above.

For the remaining two terms we have
\begin{align*}
S^\mu_{\ i}S^i_{\ \mu}v^\Lambda_+ 
&= \sum{i,\mu}[S^\mu_{\ i},S^i_{\ \mu}]v^\Lambda_+\\
&= \sum_{i,\mu}\left( S^\mu_{\ \mu} + S^i_{\ i} \right) v^\Lambda_+\\
&= m\sum_{\mu=1}^kS^\mu_{\ \mu}v^\Lambda_+ = m\sum_{\mu=1}^k\Lambda_\mu v^\Lambda_+ 
\end{align*}
and similarly
\begin{align*}
S^i_{\ \overline{\mu}} S^{\overline{\mu}}_{\ i} 
&= \sum_{i,\mu}\left( S^{\overline{\mu}}_{\ \overline{\mu}} + S^i_{\ i} \right) v^\Lambda_+\\
&= -m\sum_{\mu=1}^k\Lambda_\mu v^\Lambda_+.
\end{align*}
Thus the eigenvalue $\chi_\Lambda(I_2)$ of $I_2$ on $V(\Lambda)$ is given by
$$
\chi_\Lambda(I_2) = 2(\Lambda,\Lambda+2\rho_0) + 2m\sum_{\mu=1}^k\Lambda_\mu.
$$
By comparison with our expression for $\rho_1$ given in (\ref{rhos}), we may write
\begin{align*}
\chi_\Lambda(I_2) &= 2(\Lambda,\Lambda+2\rho_0) - 2(\Lambda,2\rho_1)\\
&= 2(\Lambda,\Lambda+2\rho)
\end{align*}
which is consistent with equation (\ref{p19star}). The important point here is that we in fact have
$$
I_2=2C_L,
$$
i.e. the second order invariant is actually {\em twice} the universal Casimir element.


\section{Characteristic identities} \label{sec9}

Let $U=U(L)$ be the universal enveloping algebra of $L=osp(m|n)$ and $\Delta:U \rightarrow U\otimes
U$ the usual co-product defined by
$$
\Delta(x)=x\otimes 1+1\otimes x,\  \ x\in L,
$$
where $1$ is the unit element of $U$. This implies that
\begin{align*}
\Delta(C_L) &= \frac12\Delta(I_2) \\
		   &= \frac12 \left( S^p_{\ q}\otimes 1 + 1\otimes S^p_{\ q} \right)(-1)^{(q)} 
\left( S^q_{\ p}\otimes 1 + 1\otimes S^q_{\ p} \right)
\end{align*}
since 
\begin{align*}
	I_2 &= S^p_{\ q}(-1)^{(q)}S^q_{\ p} = \sigma^p_{\ q}(-1)^{(q)}\sigma^q_{\ p}\\
\Rightarrow \ \ \Delta(C_L) &= \frac12\left( \sigma^p_{\ q}\otimes 1 + 1\otimes \sigma^p_{\ q} \right)
	(-1)^{(q)}\left( \sigma^q_{\ p}\otimes 1+1\otimes \sigma^q_{\ p} \right) \\
	&= C_L\otimes 1+1\otimes C_L + (-1)^{(q)}\sigma^p_{\ q}\otimes \sigma^q_{\ p}.
\end{align*}
Now let $\pi^*$ be the dual vector irreducible representation so that
$$
\pi^*(x) = -\pi^T(x),\ \ x\in L
$$
with $T$ the supertranspose. Then by definition,
\begin{align*}
	\pi^*(\sigma^p_{\ q}) &= -\left( E^p_{\ q} - (-1)^{(p)((p)+(q))}\theta_p\theta_qE^{\tilde{q}}_{\
\tilde{p}} \right)^T
\mbox{ (c.f. equation (\ref{Racahgen}) )}\\
&= -(-1)^{((p)+(q))(p)}E^q_{\ p} + \theta_p\theta_qE^{\tilde{p}}_{\ \tilde{q}}(-1)^{(p)+(q)}.
\end{align*}

We now consider the matrix
\begin{align*}
A &= -\frac12 \pi^*\otimes\id \left( \Delta(C_L) - C_L\otimes 1 - 1\otimes C_L \right) \\
  &= \frac12 \sum_{p,q}\left\{ (-1)^{((p)+(q))(p)}E^q_{\ p} -
(-1)^{(p)+(q)}\theta_p\theta_qE^{\tilde{p}}_{\ \tilde{q}}  \right\}\otimes
(-1)^{(q)}\sigma^q_{\ p}\\
&= \frac12\sum_{p,q}(-1)^{(p)(q) + (p)+(q)} E^p_{\ q}\otimes \sigma^p_{\ q} -
\frac12\sum_{p,q}\theta_{\tilde{p}}\theta_{\tilde{q}} E^p_{\ q}(-1)^{(p)}\otimes
\sigma^{\tilde{q}}_{\ \tilde{p}}\\
&= \sum_{p,q}(-1)^{(p)(q)+(p)+(q)}E^p_{\ q}\otimes \sigma^p_{\ q},
\end{align*}
where we have utilised the symmetry
$$
\sigma^{\tilde{q}}_{\ \tilde{p}} = -(-1)^{(p)((p)+(q))}\theta_p\theta_q\sigma^p_{\ q}
$$
together with 
$$
\theta_p\theta_{\tilde{p}} = (-1)^{(p)}.
$$
Thus we arrive at our characteristic matrix
$$
A = \sum_{p,q}(-1)^{(p)(q)}E^p_{\ q}\otimes \sigma^p_{\ q} (-1)^{(p)+(q)}
$$
with entries
\begin{equation}
	A_{pq} = (-1)^{(p)}\sigma^p_{\ q}.
\label{charmat}
\end{equation}
{\em Note}: The entries of this matrix are in Racah form.

It follows immediately that this matrix satisfies the polynomial identity 
$$
\prod_{p=1}^{m+n}(A-\alpha_p)=0,
$$
where the characteristic roots are given by
\begin{align*}
\alpha_p &= -\frac12\left[ \chi_{\Lambda-\varepsilon_p}(C_L) - \chi_{\delta_1}(C_L) -
\chi_{\Lambda}(C_L) \right]\\
&= -\frac12\left[ (\Lambda-\varepsilon_p,\Lambda - \varepsilon_p+2\rho) -
(\delta_1,\delta_1+2\rho)-(\Lambda,\Lambda+2\rho) \right]\\
&= \frac12 \left[ 2(\Lambda+\rho,\varepsilon_p) + m-n-1-(-1)^{(p)} \right]\\
&= (\Lambda+\rho,\varepsilon_p) + \frac12\left(m-n-1-(-1)^{(p)} \right).
\end{align*}
{\em Note}: Here we have used the fact that for the vector module $V\cong V^*$, we have
the $\mathbb{Z}$-grading
$$
V = V_0\oplus V_1\oplus V_2,
$$ 
where $V_0$ is the vector module of $gl(k)$, $V_1$ the vector module of $o(m)$ and $V_2$
the dual vector module of $gl(k)$. The weights in $V$ are the $\varepsilon_p$ ($1\leq
p\leq m+n$) and the highest weight is actually $\delta_1$.
$\blacksquare$ \medskip

Thus we have the odd characteristic roots, with $1\leq\mu\leq k$,
\begin{align*}
\alpha_\mu &= (\Lambda+\rho,\delta_\mu) + \frac12(m-n)\\
&= -\Lambda_\mu + m-n+\mu-1,\\
\alpha_{\overline{\mu}} &= -(\Lambda+\rho,\delta_\mu) + \frac12(m-n)\\
&= \Lambda_\mu - \mu+1,
\end{align*}
together with the even characteristic roots, with $1\leq i\leq h$,
\begin{align*}
\alpha_i &= (\Lambda+\rho,\varepsilon_i)+\frac12(m-n-2)\\
&= \Lambda_i+m-i-\frac{n}{2}-1,\\
\alpha_{\overline{i}} &= -(\Lambda+\rho,\varepsilon_i) + \frac12(m-n-2)\\
&= -\Lambda_i + i-\frac{n}{2} - 1,
\end{align*}
and the additional even root
$$
\alpha_{h+1} = \frac12(m-n-2)
$$
when $m=2h+1$ is odd.

If we define odd weight labels $\Lambda_\mu$, for $\mu>k$, by
$$
\Lambda_{\overline{\mu}} = m-\Lambda_\mu+1, 
$$
then the above characteristic roots can be written in the unified form
$$
\alpha_\mu = -\Lambda_\mu+m-n+\mu-1,\ \ 1\leq\mu\leq n.
$$
Similarly if we define even weight labels $\Lambda_i$, for $i>h$, according to
$$
\Lambda_{\overline{i}} = 1-\Lambda_i,
$$
which is consistent with
$$
\Lambda_{h+1}=\frac12
$$
for odd $m=2h+1$, the even characteristic roots can also be expressed in the unified form
$$
\alpha_i = \Lambda_i+m-i-\frac{n}{2}-1,\ \ 1\leq i\leq m.
$$

Thus with these conventions we arrive at the $osp(m|n)$ identity
$$
\prod_{i=1}^m(A-\alpha_i)\prod_{\mu=1}^n(A-\alpha_\mu) = 0,
$$
where we define labels $\Lambda_p$ for all $p=1,2,\ldots,m+n$ according to the above, and
our even and odd roots are
\begin{align*}
\alpha_i &= \Lambda_i+m-i-\frac{n}{2}-1,\ \ 1\leq i\leq m,\\
\alpha_\mu &= -\Lambda_\mu+m-n+\mu-1,\ \ 1\leq\mu\leq n.
\end{align*}


\section{$osp(m+1|n)\supset osp(m|n)$ branching condition} \label{branchingcondition}

Here we add an extra even index $i=0$, to give the Racah generators $\sigma^p_{\ q}$,
$0\leq p,q\leq m+n$ of $osp(m+1|n)=K$ containing the canonical subalgebra $L=osp(m|n)$. 
We have, by analogy with equation (\ref{starp17}),
the $\mathbb{Z}$-grading
$$
K = K_{-2}\oplus K_{-1}\oplus K_0\oplus K_1\oplus K_2
$$
where $K_0=o(m+1)\oplus gl(k)$ and $K_{\pm 1} = L_{\pm 1}\oplus M_{\pm}$ with
\begin{align*}
M_+ &= \mbox{span}\left\{ \sigma^\mu_{\ i=0}\ |\ 1\leq\mu\leq k\right\},\\
M_- &= \mbox{span}\left\{ \sigma^{i=0}_{\ \mu}\ |\ 1\leq \mu\leq k\right\} 
\end{align*}
and $K_{\pm 2} = L_{\pm 2}.$

We also find it convenient to set 
$$
L_+ = L_1\oplus L_2,\ \ K_+ = K_1\oplus K_2  = L_+\oplus M_+
$$
and similarly for $L_-,K_-.$

Now let $\tv(\lv)$ be an irreducible $K=osp(m+1|n)$ module with highest
$\mathbb{Z}$-graded component $\tv_0(\lv)$, constituting an irreducible $K_0$-module, such
that
$$
K_+\tv_0(\lv) = (0).
$$
On the other hand, $\tv_0(\lv)$ is an irreducible $K_0$-module and $K=K_-\oplus K_0\oplus
K_+$. By the PBW theorem  we have
\begin{align*}
\tv(\lv) &= U(K_-)\tv_0(\lv)\\
&= U(L_-)U(M_-) \tv_0(\lv).
\end{align*}
Hence setting
$
W = U(M_-)\tv_0(\lv),
$
we have
\begin{align}
\tv(\lv) = U(L_-)W. \label{starp27}
\end{align}
{\em Note}: $L_-$ and $M_-$ both transform as $L_0$-modules under the adjoint action. Also
$\tv_0(\lv)$ decomposes into irreducible $L_0$-modules according to the usual
$o(m+1)\supset o(m)$ branching rule.
$\blacksquare$ \medskip

If we set $\lv=\lv_0+\lv_1$, then we write the above decomposition into irreducible
$L_0$-modules according to
\begin{equation}
\tv_0(\lv) = \bigoplus_{\Lambda_0} V_0(\Lambda_0 + \lv_1) \label{starp28}
\end{equation} 
where the components of $\Lambda_0,$ $\lv_0$ obey the usual betweenness conditions
\begin{align}
& \lv_{i=1}\geq \Lambda_1\geq \lv_2\geq \cdots\geq \lv_h\geq\Lambda_h\geq -\lv_h,\ \
m=2h, \label{doublestarp28i}\\
& \lv_{i=1} \geq \Lambda_1\geq \lv_2\geq \cdots\geq \lv_h\geq \Lambda_h\geq |\lv_{h+1}|,\
\ m=2h+1, \label{doublestarp28ii}
\end{align}
with each such module occurring exactly once. It follows that the irreducible $L_0$-modules
occurring in $W$ have highest weights of the form 
$$
\Lambda - \delta_{\mu_1} - \delta_{\mu_2} - \cdots - \delta_{\mu_r},\ \
1\leq\mu_1\neq\mu_2\neq\cdots\neq \mu_r\leq k.
$$

We now note that every finite dimensional irreducible $K$-module admits a non-degenerate
sesquilinear form which is {\em invariant} in the sense
$$
\langle xv,w\rangle = (-1)^{(x)(v)}\langle v,x^\ddagger w\rangle
$$
for all $v,w\in \tv(\lv)$, $x\in K$, where now $\ddagger$ is a {\em super-conjugation}
operation (see Appendix A). Such a form, which is induced by an inner product on the
maximal $\mathbb{Z}$-graded component $\tv_0(\lv)$ in a natural way, has all the
properties of an inner product except it is not generally positive definite. Note that in comparison to the
$C(k+1)=osp(2|2k)$ case this inner product has a different definition. See, for example, the article
\cite{ZhaGou19902}.

From Appendix A, we note that
$$
L_{\pm}^\ddagger = L_{\mp},\ \ K_{\pm}^\ddagger = K_{\mp}.
$$
Following our $gl(m|n)$ approach \cite{GIW1} we have the following result.

\begin{lemma} \label{lemma2}
Suppose $v_+$ is an $L$-maximal weight vector in $\tv(\lv)$. Then
$$
\langle v_+,W \rangle \neq (0).
$$
\end{lemma}
\proof{
Otherwise we would have
\begin{align*}
(0) 
&= \langle U(L_+)v_+,W \rangle\\
&= \langle v_+,U(L_-)W \rangle\\
&\stackrel{(\ref{starp27})}{=} \langle v_+,\tv(\lv)\rangle
\end{align*}
$$
\Rightarrow \ \ v_+ = (0).
$$
}

Finally following exactly the same approach as for $gl(m|n)$ \cite{GIW1} we arrive at the
following {\em branching condition}. The details of the proof are given in Appendix B.

\begin{thm} \label{branchingtheorem}
Let $\tv(\lv)$ be a finite dimensional irreducible $osp(m+1|n)$ module with highest weight
$\lv$ and $v_+$ a maximal weight vector of $osp(m|n)$ of highest weight $\Lambda$. Then
\begin{itemize}
\item[(a)] $v_+$ is unique (up to scalar multiples),
\item[(b)] the components of $\Lambda$ must satisfy the betweenness conditions
(\ref{doublestarp28i})-(\ref{doublestarp28ii}) and
\begin{equation}
\lv_\mu\geq \Lambda_\mu\geq \lv_\mu -1
\ \ 1\leq\mu\leq k.
\label{braconthm1}
\end{equation}
\end{itemize}
\end{thm}

It follows that our previous approach for evaluating the reduced matrix elements and
certain Wigner coefficients based on the characteristic identities, should extend to
$osp(m|n)$.

\noindent
{\em Note}: The above also extends to $osp(1|n=2k)$ but with even index generator
$$
\sigma^{i=1}_{\ i=1}
$$
which vanishes identically. In this case the even subalgebra is simply
$L_{\overline{0}}=sp(n=2k).$


\section{$osp(m|n)$ vector operators} \label{sec11}

Recall that the $osp(m|n)$ matrix has entries given by equation (\ref{charmat}), i.e.
$A=\gamma\sigma$, with $\gamma$ the grading automorphism introduced in (\ref{gradaut}). Now let $V$
be the vector module and $\pi$ the irreducible representation of $L=osp(m|n)$ afforded by $V$. Then
a {\em vector operator} $\psi$ is equivalent to an intertwining operator on an irreducible
$L$-module $V(\Lambda)$,
$$
\psi:V\otimes V(\Lambda)\rightarrow W,
$$
with $W=$Im$\psi$ some $L$-module. Thus for $x\in L$ we have
\begin{align*}
x\psi(e_p\otimes v) &= \psi\Delta(x)(e_p\otimes v)\\
&=\psi(xe_p\otimes v) + (-1)^{(p)(x)}\psi(e_p\otimes xv)
\end{align*}
$\forall v\in V(\Lambda)$, with $\{e_p\}$ the usual basis for $V$.

Equivalently we have a collection of components $\psi^p$, defined by
$$
\psi^pv = \psi(e_p\otimes v), \ \ \forall v\in V(\Lambda)
$$
so that, for all $v\in V(\Lambda)$,
\begin{align*}
	& x\psi^p v = \psi(xe_p\otimes v) + (-1)^{(p)(x)} \psi^px v \\
	\Rightarrow & (x\psi^p - (-1)^{(p)(x)}\psi^px)v = \pi(x)_{qp}\psi^q v. 
\end{align*}
Therefore, by abstraction, we arrive at the transformation law of $osp(m|n)$ vector operators:
$$
[x,\psi^p] = \pi(x)_{qp}\psi^q,\ \ \forall x\in L,
$$
with the bracket on the left hand side being the usual graded commutator. Thus a vector operator
satisfies
$$
[\sigma^p_{\ q},\psi^r] = \pi(\sigma^p_{\ q})_{sr} \psi^s.
$$
Using
$$
\pi(\sigma^p_{\ q}) = e_{pq} - (-1)^{(p)((p)+(q))}\theta_p\theta_qe_{\tilde{q}\tilde{p}},
$$
where now $e_{pq}$ is an elementary matrix, it follows that
\begin{align*}
\pi(\sigma^p_{\ q})_{sr} &= \delta_{qr}\delta^p_{\ s} -
	(-1)^{(p)((p)+(q))}\theta_p\theta_q\delta_{\tilde{q}s}\delta_{\tilde{p}r} \\
\Rightarrow \ \ [\sigma^p_{\ q},\psi^r] &= \delta^r_{\ q}\psi^p -
	(-1)^{(p)((p)+(q))}\theta_p\theta_q\delta^{\tilde{p}}_{\ r}\psi^{\tilde{q}}.
\end{align*}
this is the transformation law for a vector operator.

Similarly we say that the operators $\phi_p$ transform as a {\em contragredient vector operator} if
\begin{align*}
[\sigma^p_{\ q},\phi_r] &= \pi^*(\sigma^p_{\ q})_{sr}\phi_s \\
&= -\pi^T(\sigma^p_{\ q})_{sr}\phi_s \\
&= -\left\{ e_{pq}^T - (-1)^{(p)((p)+(q))} \theta_p\theta_qe_{\tilde{q}\tilde{p}}^T
\right\}_{sr}\phi_s\\
&= -(-1)^{(p)((p)+(q))}\left\{ \delta^p_{\ r}\phi_q - (-1)^{((p)+(q))(q)}\delta^{\tilde{q}}_{\
r}\phi_{\tilde{p}} \right\}.
\end{align*}

\ \\

\noindent
\underline{Motivating example}: $osp(m+1|n)\supset osp(m|n)$

\ \\

Here we have the additional index $i=0$ giving rise to the operators
$$
\psi^p = \sigma^p_{\ 0},\ \ \phi_p = \sigma^0_{\ p}\ \ (1\leq p\leq m+n)
$$
which are easily seen to transform as vector (respectively contragredient vector) operators with
respect to $osp(m|n)$.


\section{Characteristic identities and reduced matrix elements} \label{sec12}

We now let 
$$
A^p_{\ q} = (-1)^{(p)}\sigma^p_{\ q},\ \ 1\leq p,q\leq m+n
$$
be the $osp(m|n)$ matrix with characteristic roots $\alpha_r$ and 
$$
B^p_{\ q} = (-1)^{(p)}\sigma^p_{\ q},\ \ 0\leq p,q\leq m+n
$$
the corresponding $osp(m+1|n)$ matrix with characteristic roots $\beta_r$ ($0\leq r\leq m+n$). We
have also the associated projections
\begin{align}
P[r] &= \prod_{q\neq r,q=1}^{m+n} \left( \frac{A-\alpha_q}{\alpha_r-\alpha_q} \right),\label{projA} \\
Q[r] &= \prod_{q\neq r,q=0}^{m+n} \left( \frac{B-\beta_q}{\beta_r-\beta_q} \right), \label{projB}
\end{align}
which satisfy
$$
P[q]P[r] = \delta_{qr}P[q],\ \ Q[q]Q[r] = \delta_{qr}Q[q]
$$
and the identity resolutions
$$
\sum_{q=1}^{m+n}P[q]_{rs} = \delta_{rs},\ \ 
\sum_{q=0}^{m+n}Q[q]_{rs} = \delta_{rs}. 
$$
{\em Remark}: 
As we have pointed out in our previous work relating to $gl(m|n)$ \cite{GIW1}, there may exist
irreducible representations for which the characteristic roots coincide. In such a case, the
projections defined above in (\ref{projA}) or (\ref{projB}) would be undefined. To circumvent this
issue, as in the case of $gl(m|n)$, we point out that the set of highest weights to which such
characteristic roots correspond is closed in the Zariski topology \cite{Humphreys} on $H^*$, the Cartan subalgebra
dual. It follows that there is a dense subset of $H^*$ on which the characteristic roots are distinct.
Without loss of generality, we therefore make the assumption that all characterstic roots are
distinct. This means that in practice, when applying the final formulae for the invariants which
appear as rational polynomial
functions of the characteristic roots, we must always be mindful of the fact that we are relying on
analytic continuation to define the function in certain pathological cases, and that terms
in the numerator and denominator should be cancelled where required. 
This can always be done since polynomial functions on $H^*$ are continuous in the Zariski topology.
$\blacksquare$ \medskip

Following our $gl(m|n)$ derivation \cite{GIW1} we have for $1\leq r\leq m+n$,
$$
B^r_{\ s} Q[q]^s_{\ 0} + B^r_{\ 0}C_q = \beta_q Q[q]^r_{\ 0},
$$
where $C_q$ denotes the $osp(m|n)$ invariant
$$
C_q = Q[q]^0_{ \ 0}
$$
which has eigenvalues that determine the squares of a certain Wigner coefficient.

\noindent
{\em Note}: Recall that for $osp(m|n)$,
\begin{align*}
\alpha_r &= -\frac12 \left[ \chi_{\Lambda-\varepsilon_r}(C_L) - \chi_{\delta_1}(C_L) -
\chi_\Lambda(C_L) \right]\\
&= (\Lambda+\rho,\varepsilon_r)+\frac12 (m-n-1-(-1)^{(r)}),
\end{align*}
with a similar expression for $\beta_r$. The important point is that $P[r],$ $Q[r]$ project onto
a submodule with highest weight $\Lambda-\varepsilon_r$ in the tensor product module $V\otimes
V(\Lambda)$ (with $V$ the vector module). 
$\blacksquare$ \medskip

Thus rearranging we may write 
\begin{equation}
	(-1)^{(r)}\psi^rC_q = (\beta_q-A)^r_{\ s}Q[q]^s_{\ 0},
\label{starp4II}
\end{equation}
where
$$
\psi^r = (-1)^{(r)}B^r_{\ 0} = \sigma^r_{\ 0},\ \ 1\leq r\leq m+n
$$
is a vector operator of $osp(m|n)$ as we have seen.

At this point we note degeneracies between the odd roots of $osp(m+1|n)$ and those of $osp(m|n)$
which parallels the situation with $gl(m|n)$ (but with even roots in that case). In terms of the
$osp(m+1|n)$ representation labels $\lv_\mu$ ($1\leq \mu\leq k$) we have the characteristic roots
$$
\beta_\mu = -\lv_\mu + m-1-n+\mu-1,\ \ 1\leq \mu\leq n,
$$
where we define the representation labels $\lv_\mu$ for $\mu>k=n/2$ by
$$
\lv_{\overline{\mu}} = m-\lv_\mu+1
$$
as before. If $\Lambda_\mu$ are the $osp(m|n)$ representation labels so
$$
\alpha_\mu = -\Lambda_\mu + m-n+\mu-1,\ \ 1\leq \mu\leq n,
$$
we have from the branching condition (\ref{braconthm1}) of Theorem \ref{branchingtheorem} that 
$$
\beta_\mu = 
\left\{  
\begin{array}{rl}
1+\alpha_\mu, & \Lambda_\mu = \lv_\mu,\\
\alpha_\mu, & \Lambda_\mu = \lv_\mu - 1,
\end{array}
\right.
$$
while
$$
\beta_{\overline{\mu}} = \lv_\mu - \mu - \frac12 =  
\left\{  
\begin{array}{rl}
\alpha_{\overline{\mu}}, & \Lambda_\mu = \lv_\mu,\\
1+\alpha_{\overline{\mu}}, & \Lambda_\mu = \lv_\mu - 1.
\end{array}
\right.
$$
We now introduce the odd index set $I_1$ as follows. For $1\leq \mu\leq k$ we take
$$
\mu\in I_1 \ \ \Leftrightarrow\ \ \lv_\mu = \Lambda_\mu+1 \ (\mbox{or }\beta_\mu=\alpha_\mu),
$$
otherwise we take $\overline{\mu}\in I_1$, i.e.
$$
\overline{\mu}\in I_1\ \ \Leftrightarrow\ \ \lv_\mu = \Lambda_\mu.
$$
We define $\overline{I}_1$ by
$$
\overline{I}_1 = \{ 1\leq\mu\leq n\ |\ \mu \notin I_1\},
$$
so for $1\leq \mu\leq k$ we have 
\begin{align*}
\mu\in\overline{I}_1 \mbox{ if } \lv_\mu = \Lambda_\mu,\\
\overline{\mu}\in\overline{I}_1 \mbox{ if } \lv_\mu = 1+\Lambda_\mu.
\end{align*}
Therefore, for $1\leq\mu\leq k,$
$$
\mu\in\overline{I}_1\ \ \Rightarrow \\ C_{\mu}=0
$$
since $\Lambda_\mu=\lv_\mu$ already takes its maximum value, whilst if $\mu\in I_1$ we have
$C_{\overline{\mu}}=0.$ Thus for every odd index, $\mu$, we have
$$
C_\mu=0, \mbox{ if } \mu\in\overline{I}_1
$$ 
and
$$
C_{\overline{\mu}} = 0, \mbox{ if } \mu\in I_1.
$$
{\em Note}: As for $gl(m|n)$, we have also the full index sets
$$
I = I_0\cup I_1,\ \ \tilde{I} = I\cup \{i=0\},
$$
where $I_0 = \{ i=1,2,\ldots,m\}$ is the set of even indices.
$\blacksquare$ \medskip

We shall also consider the contragredient vector operator
$$
\phi_r = \sigma^0_{\ r},\ \ 1\leq r\leq m+n.
$$
Then $\phi_r,$ $\psi^r$ may be resolved into shift components in the usual way:
$$
\phi_r = \sum_{q=1}^{m+n}\phi[q]_r,\ \ \psi^r = \sum_{q=1}^{m+n}\psi[q]^r
$$
where 
\begin{align}
	\psi[q]^r &= (-1)^{(r)} P[q]^r_{\ s} (-1)^{(s)}\psi^s, \label{starp7IIi}\\
\phi[q]_r &= \phi_sP[q]^s_{\ r}, \label{starp7IIii}
\end{align}
with $P[q]$ the projection introduced earlier. Then we clearly have that if $q$ is an odd index then
$$
\psi[q]=0 \mbox{ if } q\in \overline{I}_1,
$$
while
$$
\phi[q]=0 \mbox{ if } q\in I_1.
$$
{\em Note}: Some care needs to be taken here. Actually $\phi[q]_r$ will {\em decrease} $\Lambda$ by
$-\varepsilon_q$, i.e. $\phi[q]$ affects the shift $\Lambda \rightarrow \Lambda-\varepsilon_q$.
Therefore for $q=\mu$ an odd index, $1\leq \mu\leq k,$ we see that $\phi[\mu]$ affects the shift
$\Lambda\rightarrow \Lambda-\delta_\mu$ so that $\phi[\mu]$ will {\em decrease} label $\Lambda_\mu$
by 1 unit. If $\Lambda_\mu = \lv_\mu - 1,$ i.e. $\mu\in I_1,$ then $\phi[\mu]$ must  vanish, and
vice-versa for $\mu\in\overline{I}_1$.
$\blacksquare$ \medskip

Inverting equation (\ref{starp4II}) and resolving $\psi^r$ into its shift components allows us to
write 
\begin{equation}
	Q[q]^r_{\ 0} = (-1)^{(r)}\sum_{s\in I} (\beta_q-\alpha_s)^{-1}\psi[s]^rC_q,\ \ q\in \tilde{I},
\label{aboveequ}
\end{equation}
where we have used the fact that $\psi[s]^r,$ $Q[q]^r_{\ 0}$ and $C_q$ all vanish for
$s,q\in\overline{I}_1$. Using the easily established relation
$$
(\beta_q-\alpha_s)^{-1}\psi[s] = \psi[s](\beta_q-\alpha_s-(-1)^{(s)})^{-1}
$$
except for odd $m=2h+1$ where we have
$$
(\beta_q-\alpha_i)^{-1}\psi[i] = \psi[i](\beta_q - \alpha_i)^{-1},\ \ i=h+1,
$$
equation (\ref{aboveequ}) may be expressed in the form
\begin{equation}
	Q[q]^r_{\ 0} = (-1)^{(r)}\sum_{s\in I} \psi[s]^r (\beta_q-\alpha_s-(-1)^{(s)} + \delta_{s,h+1})^{-1}C_q,
\label{starp8II}
\end{equation}
where the term $\delta_{s,h+1}$ only contributes when $m=2h+1$ is odd.

\noindent
{\em Note}: Using the relation
$$
\left( \sigma^p_{\ q} \right)^\ddagger = (-1)^{((p)+(q))(p)}\sigma^q_{\ p}
$$
we have the easily established relations
$$
\left( \psi^p \right)^\ddagger = (-1)^{(p)}\phi_p,\ \ \phi_p^\ddagger = \psi^p.
$$
Thus resolving into shift components
$$
\Rightarrow \ \ \sum_r\phi[r]_p^\ddagger = \sum_r\psi[r]^p
$$
so we arrive at
$$
\phi[r]_p^\ddagger = \psi[r]^p.
$$
This can also be seen by noting for our $osp(m|n)$ matrix $A$ that
$$
A_{pq}^\ddagger = (-1)^{(q)((p)+(q))}A_{qp}
$$
and more generally or matrix powers 
$$
\left( A_{pq}^N \right)^\ddagger = (-1)^{((p)+(q))(q)}A_{qp}^N.
$$

The matrix $A$ acts naturally on the right of $\phi_r = B^0_{\ r} = \sigma^0_{\ r}$ so that
\begin{align*}
\left( \phi[q]_r \right)^\ddagger &= \left( \phi_s P[q]_{sr} \right)^\ddagger \\
&= (-1)^{(s)((s)+(r))}\left( P[q]_{sr} \right)^\ddagger \phi_s^\ddagger \\
&= (-1)^{(s)((s)+(r))}(-1)^{((r)+(s))(r)} P[q]_{rs}\psi^s \\
&= (-1)^{(r)}P[q]_{rs}(-1)^{(s)}\psi^s \\
&\stackrel{(\ref{starp7IIi})}{=} \psi[q]^r
\end{align*}
as expected.

\noindent
{\em Note}: For odd index $p=\mu$, $1\leq\mu\leq k$, or even index $p=i$, $1\leq i\leq h$,
$\phi[p]_r$ {\em decreases} the representation label $\Lambda_p$ by 1 unit so $\psi[p]^r$ must
{\em increase} it.
$\blacksquare$ \medskip

Our analysis now follows closely that of $gl(m|n)$ \cite{GIW1}. Summing equation (\ref{starp8II})
over $q$ gives
$$
(-1)^{(r)}\sum_{s\in I}\psi[s]^r\sum_{q\in\tilde{I}}(\beta_q-\alpha_s-(-1)^{(s)} +
\delta_{s,h+1})^{-1} C_q = 0
$$
\begin{equation}
\Rightarrow \ \ \sum_{q\in\tilde{I}} (\beta_q-\alpha_s - (-1)^{(s)}+\delta_{s,h+1}) C_q=0,\ \ s\in
I.
\label{starp10II}
\end{equation}
Also using the resolution
\begin{equation}
\sum_{q\in\tilde{I}} C_q = 1,
\label{doublestarp10II}
\end{equation}
we obtain $|I|+1$ equations in $|\tilde{I}| = |I|+1$ unknowns which uniquely determines the
$osp(m|n)$ invariants $C_q$. By this means we arrive at the formula
\begin{equation}
C_q = \prod_{k\neq q}^{\tilde{I}} (\beta_q-\beta_k)^{-1}\prod_{r\in I}
(\beta_q-\alpha_r-(-1)^{(r)}+\delta_{r,h+1}).
\label{triplestarp10II}
\end{equation}
{\em Note}: Again, the term $\delta_{r,h+1}$ only applies when $m=2h+1$ is odd. Thus, to be precise, we have the
following formulae:
$$
C_q = \left\{  
\begin{array}{rl}
\displaystyle{\prod_{k\neq q}^{\tilde{I}} (\beta_q-\beta_k)^{-1}\prod_{r\in I}(\beta_q -
\alpha_r-(-1)^{(r)})}, &
m=2h,\\
\displaystyle{\prod_{k\neq q}^{\tilde{I}} (\beta_q-\beta_k)^{-1}\prod_{r\in I}(\beta_q -
\alpha_r-(-1)^{(r)}+\delta_{r,h+1})}, & m=2h+1.
\end{array}
\right.
$$
However, it is convenient to use the unified formula of equation (\ref{triplestarp10II}).
$\blacksquare$ \medskip

We now invert equation (\ref{starp8II}) by writing
\begin{equation}
	\psi[p]^r = (-1)^{(r)}\sum_{s\in\tilde{I}} Q[s]^r_{\ 0} \gamma_{ps},\ \ p\in I
\label{starp11II}
\end{equation}
for suitable coefficients $\gamma_{ps},$ $p\in I$, $s\in \tilde{I}$. This leads us to consider the
unique solutions $\gamma_{ps}$ to the set of equations
\begin{align}
	\sum_{s\in\tilde{I}}\gamma_{ps}(\beta_s - \alpha_r - (-1)^{(r)}+\delta_{r,h+1})^{-1}C_s &=
\delta_{pr} \label{doublestar1p11II}\\
\sum_{s\in\tilde{I}}\gamma_{ps} C_s &=0,\ \ p,r\in I. \label{doublestar2p11II}
\end{align}
Then for each $p\in I$ this yields $|\tilde{I}| = |I|+1$ equations in $|\tilde{I}|$ unknowns
$\gamma_{ps}$, $s\in \tilde{I}.$

\noindent
{\em Remark}: Since then we obtain
\begin{align*}
\sum_{s\in\tilde{I}} Q[s]^r_{\ 0}\gamma_{ps}
& \stackrel{(\ref{starp8II})}{=} (-1)^{(s)}\sum_{s\in \tilde{I}}\sum_{q\in I}\psi[q]^r
(\beta_s - \alpha_q-(-1)^{(q)} + \delta_{q,h+1})^{-1}C_s\gamma_{ps} \\
& \stackrel{(\ref{doublestar1p11II})}{=} (-1)^{(r)} \sum_{q\in I} \psi[q]^r
\delta_{pq}\\
& = (-1)^{(r)}\psi[p]^r
\end{align*}
as required.
$\blacksquare$ \medskip

The above equations (\ref{doublestar1p11II}) and (\ref{doublestar2p11II}) are easily solved using
matrix methods and yield the unique solution
$$
\gamma_{ps} = \gamma_p\left( \beta_s - \alpha_p - (-1)^{(p)} + \delta_{p,h+1} \right)^{-1},\ \ p\in
I,\ s\in\tilde{I}
$$
where
\begin{align}
	\gamma_p &= (-1)^{|I|} \prod_{r\neq p}^I\left( \alpha_p-\alpha_r+(-1)^{(p)} - (-1)^{(r)} +
\delta_{r,h+1} - \delta_{p,h+1} \right)^{-1}
\prod_{q\in \tilde{I}}\left( \beta_q - \alpha_p - (-1)^{(p)} + \delta_{p,h+1} \right).
\label{starp12II}
\end{align}
As for the case of $gl(m|n)$ these invariants have a natural interpretation. From the remarks above
we have
$$
\psi[p]^r = (-1)^{(r)} \sum_{s\in\tilde{I}} Q[s]^r_{\ 0}\gamma_{ps}
$$
so that (summation over repeated indices here and below)
\begin{align*}
	\phi[p]_{r} (-1)^{(r)} \psi[p]^r &= \phi_r (-1)^{(r)}
\psi[p]^r\\
&= \phi_r\sum_{s\in \tilde{I}}Q[s]^r_{\ 0} \gamma_{ps} \\
&= \sum_{s\in \tilde{I}} B^0_{\ r} Q[s]^r_{\ 0} \gamma_{ps} \\
&= \sum_{s\in\tilde{I}}(\beta_s - B^0_{\ 0})C_s\gamma_{ps}
\end{align*}
where we have utilised the $osp(m+1,n)$ characteristic identity. Also using 
$$
B^{i=0}_{\ i=0} = \sigma^0_{\ 0} = 0,
$$
we obtain
\begin{align*}
	\phi[p]_{r}(-1)^{(r)}\psi[p]^r &= \sum_{s\in \tilde{I}}\beta_s
C_s\gamma_{ps}\\
& \stackrel{(\ref{starp12II})}{=} \gamma_p\sum_{s\in\tilde{I}}\beta_sC_s
\left( \beta_s - \alpha_p - (-1)^{(p)}+\delta_{p,h+1} \right)^{-1}\\
&= \gamma_p\left\{ \sum_{s\in\tilde{I}}C_s + (\alpha_p + (-1)^{(p)}-
\delta_{p,h+1})\sum_{s\in\tilde{I}} C_s\left( \beta_s - \alpha_p - (-1)^{(p)} + \delta_{p,h+1}
\right)^{-1} \right\} \\
& \stackrel{(\ref{starp10II}),(\ref{doublestarp10II})}{=} \gamma_p,\ \ p\in I.
\end{align*}
{\em Remarks}: It follows that the invariants $\gamma_p$, given explicitly by formula
(\ref{starp12II}) above, are closely related to the {\em reduced matrix elements} of the vector
operator $\psi^r = \sigma^r_{\ 0}$ (c.f. the case of $gl(m|n)$ in \cite{GIW1}).


\section{Reduced matrix elements and reduced Wigner coefficients} \label{sec13}

Since $P[r]$ and $Q[r]$ both determine projections, it can be shown that
\begin{equation}
	(P[r]Q[s]P[r])_{p q} = \lambda P[r]_{pq},\ \ 1\leq
	p,q\leq m+n
	\label{starp13II}
\end{equation}
for some $osp(m|n)$-invariant $\lambda$ commuting with both $Q[s]$ and $P[r]$ (generalised
angle operator). On the other hand from the $osp(m+1|n)$ characteristic identity we have 
$$
(BQ[s])^p_{\ q} = \beta_sQ[s]^p_{\ q},\ \ 0\leq p,q\leq m+n
$$
which may be expressed 
$$
B^p_{\ 0}Q[s]^0_{\ q} = (\beta_s - A)^p_{r}Q[s]^r_q
$$
(summation on $r$ from 1 to $m+n$), or
$$
(-1)^{(p)}\psi^p Q[s]^0_{\ q} = (\beta_s - A)^p_{\ r}Q[s]^r_{\ q}.
$$
Multiplication on the left by $P[t]$ we obtain in view of equation (\ref{starp7IIi}) and
the $osp(m|n)$ characteristic identity
$$
\psi[t]^p Q[s]^0_{\ q} = (-1)^{(p)}(\beta_s - \alpha_t) P[t]^p_{\
r} Q[s]^r_{\ q},
$$
which may be rearranged to give
\begin{equation}
	(P[t]Q[s])^p_{\ q} = (-1)^{(p)}(\beta_s -
	\alpha_t)^{-1}\psi[t]^p Q[s]^0_{\ q} \label{starp14II}
\end{equation}
with $1\leq p\leq m+n$, $0\leq q\leq m+n$, $t\in I$, $s\in \tilde{I}$.

On the other hand the $osp(m+1|n)$ characteristic identity also implies 
$$
(Q[s]B)^0_{\ q} = Q[s]^0_{\ q}\beta_s
$$
or
\begin{align*}
	Q[s]^0_{\ 0} B^0_{ \ q} &= Q[s]^0_{\ r}(\beta_s - A)^r_{\ q} \\
	\Rightarrow \ \ C_s \phi_s &= Q[s]^0_{ \ r}(\beta_s - A)^r_{\ q},\ \
	1\leq q\leq m+n.
\end{align*}
In this case we multiply on the right by $P[t]$ to give
\begin{align}
	C_s\phi[t]_r &= Q[s]^0_{\ r} (\beta_s - \alpha_t) P[t]^r_{\ q}
	\nonumber\\
	\Rightarrow \ \ (Q[s]P[t])^0_{\ r} &= C_s \phi[t]_q(\beta_s -
	\alpha_t)^{-1} \nonumber \\
	&= C_s(\beta_s-\alpha_t - (-1)^{(t)} + \delta_{t,h+1})^{-1} \phi[t]_q,
	\label{doublestarp14II}
\end{align}
for $t\in I,$ $s\in\tilde{I}$.

Thus multiplying equation (\ref{starp14II}) on the right by $P[t]$ gives
\begin{align}
	(P[t]Q[s]P[t])^p_{\ q} &= (-1)^{(p)}(\beta_s -
\alpha_t)^{-1}\psi[t]^p (Q[s]P[t])^0_{\ q} \nonumber \\
&\stackrel{(\ref{doublestarp14II})}{=} (-1)^{(p)} (\beta_s - \alpha_t)^{-1}
\psi[t]^p C_s (\beta_s - \alpha_t - (-1)^{(t)}+\delta_{t,h+1})^{-1} \phi[t]_q.
\label{triplestarp14II}
\end{align}
{\em Note}: Since $C_s$ is expressible in terms of the $\beta_q$, $\alpha_r$ in accordance
with equation (\ref{triplestarp10II}), it follows that we must have
$$
(P[t]Q[s]P[t])^p_{\ q} = \lambda' \psi[t]^p\phi[t]_q(-1)^{(p)}
$$
for some function $\lambda'$ of the $\beta_q$, $\alpha_r$. 
$\blacksquare$ \medskip

It then follows from equation
(\ref{starp13II}) that we may write
\begin{equation}
	(-1)^{(p)}\psi[t]^p\phi[t]_q = \mu_tP[t]_{pq}, \ \ 1\leq
	p,q\leq m+n \label{starp15II}
\end{equation}
where $\mu_t$ is some invariant. On the other hand we note that 
$$
X_{pq} \equiv (-1)^{(p)} \psi[t]^p\gamma_t^{-1}\phi[t]_q = \nu
P[t]_{pq}
$$
with $\nu$ some function of the $\beta_t$, $\alpha_t$ (c.f. equation (\ref{starp12II}). Since
$$
\gamma_t = \phi[t]_\beta(-1)^{(q)}\psi[t]^q,\ \ t\in I
$$
we notice that $X$ is {\em idempotent}, i.e. $X^2=X$, which implies that $\nu^2=\nu$ or
$\nu(\nu-1)=0$. Thus for a non-zero contribution we must have $\nu=1$ which gives
\begin{equation}
	(-1)^{(p)}\psi[t]^p \gamma_t^{-1}\phi[t]_q = P[t]_{pq}.
	\label{doublestarp15II}
\end{equation}
By comparison with equation (\ref{starp15II}) above, we must have
$$
\psi[t]^p\gamma_t^{-1} = \mu_t^{-1} \psi[t]^p
$$
from which we obtain
\begin{equation}
	\mu_t = (-1)^{|I|}\prod_{s\neq t\in I}(\alpha_t - \alpha_s - (-1)^{(s)} + \delta_{s,h+1} -
	\delta_{s,\overline{t}})^{-1}\prod_{r\in\tilde{I}}(\beta_r - \alpha_t).
	\label{triplestarp15II}
\end{equation}
{\em Note}: It is understood that the term $\delta_{s,\overline{t}}$ only contributes when $t=i$ is
{\em even}. When $t$ is odd and $t\in I$, clearly $\overline{t}\notin I$ so this term vanishes.
$\blacksquare$ \medskip

Now returning to equation (\ref{triplestarp14II}), we note, from formula (\ref{triplestarp10II})
that
\begin{align*}
	\psi[t]^p C_s (\beta_s - \alpha_t - (-1)^{(t)}+\delta_{t,h+1})^{-1} 
	=&C_s(\beta_s-\alpha_t - (-1)^{(t)}+\delta_{t,h+1})^{-1}(\beta_s-\alpha_{\overline{t}}
	- (-1)^{(t)} + \delta_{t,h+1})^{-1} \\
	& \ \ \times (\beta_s-\alpha_{\overline{t}} - 2(-1)^{(t)} +
	2\delta_{t,h+1})\psi[t]^p,\ \ t\in I
\end{align*}
where it is understood that the last two terms only occur if also $\overline{t}\in I$. In other
words they only occur for $t=i$ even. 

\noindent
{\em Remark}: Thus for $t=\mu$ odd we have
$$
\psi[\mu]^p C_s(\beta_s - \alpha_\mu+1)^{-1} = C_s(\beta_s-\alpha_\mu+1)^{-1}\psi[\mu]^p
$$
but for $t=i$ even we have
\begin{align*}
	\psi[i]^p C_s (\beta_s - \alpha_i - 1+\delta_{i,h+1})^{-1} 
	=&C_s(\beta_s-\alpha_i - 1+\delta_{i,h+1})^{-1}(\beta_s-\alpha_{\overline{i}}
	- 1 + \delta_{i,h+1})^{-1} \\
	& \ \ \times (\beta_s-\alpha_{\overline{i}} - 2 +
	2\delta_{i,h+1})\psi[i]^p.
\end{align*}
These two cases are summarised above.
$\blacksquare$ \medskip

Thus substituting into equation (\ref{triplestarp14II}) we arrive at
\begin{align*}
	(P[t]Q[s]P[t])^p_{\ q} =&
	(-1)^{(p)}(\beta_s-\alpha_t)^{-1}(\beta_s-\alpha_t-(-1)^{(t)}+\delta_{t,h+1})^{-1}
	(\beta_s - \alpha_{\overline{t}} - (-1)^{(t)} + \delta_{t,h+1})^{-1} \\
	&\ \ \times
	(\beta_s -\alpha_{\overline{t}} - 2(-1)^{(t)} + 2\delta_{t,h+1})  C_s  \psi[t]^p  \phi[t]_q.
\end{align*}
Finally, using equation (\ref{starp15II}), we arrive at 
\begin{align}
	(P[t]Q[s]P[t])^p_{\ q} =&
	(-1)^{(p)}(\beta_s-\alpha_t)^{-1}(\beta_s-\alpha_t-(-1)^{(t)}+\delta_{t,h+1})^{-1}
	(\beta_s - \alpha_{\overline{t}} - (-1)^{(t)} + \delta_{t,h+1})^{-1} \nonumber\\
	&\ \ \times
	(\beta_s -\alpha_{\overline{t}} - 2(-1)^{(t)} + 2\delta_{t,h+1})  C_s  \mu_t P[t]^p_{\
	q}. 
	\label{starp17II}
\end{align}
Thus we may write (c.f. equation (\ref{starp13II}))
\begin{equation}
	(P[t]Q[s]P[t])^p_{\ q} = \omega_{s,t}P[t]^p_{\ q}
	\label{doublestarp17II}
\end{equation}
where 
\begin{align}
	\omega_{s,t} =& C_s \mu_t
	(\beta_s-\alpha_t)^{-1}(\beta_s-\alpha_t-(-1)^{(t)}+\delta_{t,h+1})^{-1}
	(\beta_s - \alpha_{\overline{t}} - (-1)^{(t)} + \delta_{t,h+1})^{-1} \nonumber\\
	&\ \ \times
	(\beta_s -\alpha_{\overline{t}} - 2(-1)^{(t)} + 2\delta_{t,h+1}). 
	\label{triplestarp17II}
\end{align}

As for $gl(m|n)$ \cite{GIW1}, the invariants $\mu_t$ of equation (\ref{triplestarp15II}) determine squares of reduced matrix elements,
while the invariants $C_k$, $\omega_{s,t}$ of equations (\ref{triplestarp10II}),
(\ref{triplestarp17II}) determine squared reduced Wigner coefficients. Together they determine
matrix elements of the generators as for $gl(m|n)$.
This problem will be investigated in a future article. Of particular interest is the determination
of a full branching rule, considering the
occurrence of indecomposable representations, and interpreting the matrix element formulae that arise
in these situations.


%
%
%
\section*{Acknowledgements}

This work was supported by the Australian Research Council through Discovery Project DP140101492.

\appendix


\section*{Appendix A: super-conjugation}

Recall the definition of {\em graded transpose} $T$ for an $(m+n)\times(m+n)$ matrix $X$:
$$
X^T_{pq} = (-1)^{(X)(q)} X_{qp}.
$$
This leads to the super adjoint $\ddagger$ defined by
$$
X^\ddagger_{pq} = \left( \overline{X}^T \right)_{pq}
$$
where as usual the over bar denotes complex conjugation. We note the following property of
the super adjoint:
$$
(XY)^\ddagger = (-1)^{(X)(Y)} Y^\ddagger X^\ddagger
$$
for homogeneous $X,Y$. In particular, if $E_{pq}$ is an elementary matrix defining the
Lie superalgebra $gl(m|n)$, we have
$$
E^\ddagger_{pq} = E_{pq}^T = (-1)^{((p)+(q))(p)}E_{qp}.
$$
Thus by abstraction we have a super adjoint operation $\ddagger$ on the Lie superalgebra
$gl(m|n)$ defined by
$$
e_{pq}^\ddagger = (-1)^{((p)+(q))(p)}e_{qp}
$$
which is easily seen to be consistent with the graded commutation relations. 

This in turn induces a super adjoint operation $\ddagger$ on $osp(m|n)$ defined by
\begin{align*}
\left( \sigma^p_{\ q} \right)^\ddagger &= e^\ddagger_{pq} -
	(-1)^{(p)((p)+(q))}\theta_p\theta_q  e^\ddagger_{\tilde{q}\tilde{p}}\\
	&= (-1)^{(p)((p)+(q))} e_{qp} - \theta_p\theta_q e_{\tilde{p}\tilde{q}} (-1)^{(p)+(q)}\\
 &= (-1)^{(p)((p)+(q))} \left( e_{qp} - (-1)^{((p)+(q))(q)} \theta_p\theta_q
e_{\tilde{p}\tilde{q}} \right)\\
&= (-1)^{(p)((p)+(q))}\sigma^q_{\ p}
\end{align*}
which is also consistent with the graded commutation relations. The important point is
that $osp(m|n)$ is stable under $\ddagger$, i.e. it admits a super-conjugation operation.

\noindent
{\em Note}: $gl(m|n)$ also admits a normal conjugation operation defined by
$$
e^\dagger_{pq} = e_{qp}
$$
which is also consistent with the $gl(m|n)$ graded commutation relations. Unfortunately
$osp(m|n)$ is not stable under $\dagger$ which is consistent with the fact that a Type II
basic classical Lie superalgebra does not admit a conjugation operation. We do, however,
see that it admits a super-conjugation operation.
$\blacksquare$ \medskip

Finally, given an inner product on the maximal $\mathbb{Z}$-graded component
$V_0(\Lambda)$ of an irreducible $L$-module $V(\Lambda)$, we have a naturally induced form
$\langle ~,~\rangle$ on all of $V(\Lambda)$ satisfying
$$
\langle av,w\rangle = (-1)^{(a)(v)}\langle v,a^\ddagger w\rangle
$$
for all homogeneous $a\in L$, $v\in V(\Lambda)$ (c.f. Gould and Zhang \cite{ZhaGou1990}).

It is also worth noting that for the Cartan-Weyl generators
$$
S^p_{\ q} = \left(M^{-1}\right)^p_{\ p'}\sigma^{p'}_{\ q'} M^{q'}_{\ q}
$$
we have 
$$
\left( S^p_{\ q} \right)^\ddagger = \left( \overline{M}^{-1} \right)^p_{\ p'}
\left( \sigma^{p'}_{\ q'} \right)^\ddagger \overline{M}^{q'}_{\ q}
$$
where we have used the anti-linear property of $\ddagger$, i.e.
$$
(\alpha X)^\ddagger = \overline{\alpha}X^\ddagger,\ \ \forall \alpha\in\mathbb{C}.
$$
Therefore,
\begin{align*}
	(S^p_{\ q})^\ddagger &= {\left( M^\dagger\right)^{-1}}^{p'}_{\ p}(-1)^{((p)+(q))(p)}\sigma^{q'}_{\
p'} \left( M^\dagger \right)^q_{\ q'} \\
&= (-1)^({(p)+(q))(p)}M^{p'}_{\ p}\sigma^{q'}_{\ p'}\left( M^{-1} \right)^q_{\ q'}\\
&= (-1)^{((p)+(q))(p)}\left( M^{-1} \right)^q_{\ q'}\sigma^{q'}_{\ p'}M^{p'}_{\ p}\\
&=   (-1)^{((p)+(q))(p)} S^q_{\ p}.
\end{align*}
It follows immediately that
$$
L^\ddagger_{\pm} = L_{\mp},\ \ L^\ddagger_0 = L_0
$$
as we wished to show.


\section*{Appendix B: Branching condition}

Here we give a careful proof of the branching condition of Theorem \ref{branchingtheorem}.
Throughout we adopt our previous notation and set $K = osp(m+1|n) \supset L = osp(m|n),$
\begin{align*}
	K_+ &= K_1\oplus K_2\\
	    &= L_+ \oplus M_+,\ \ M_+ = \mbox{span}\left\{ \sigma^\mu_{\ i=0}\ |\ 1\leq
\mu\leq k \right\}
\end{align*}
and similarly for $K_-$. We let $\tv(\lv)$ be a finite dimensional irreducible $K$-module
with highest weight
\begin{align*}
	\lv &= \lv^{(0)} + \lv^{(1)}\\
	    &= \sum_{i=1}^h \lv_i\varepsilon_i + \sum_{\mu=1}^k \lv_\mu\delta_\mu
\end{align*}
and we let $\tv_0(\lv)$ be the maximal $\mathbb{Z}$-graded component which constitutes an
irreducible $K_0 = o(m+1)\oplus gl(k)$ module.

Following our previous approach we set
$$
W = U(M_-)\tv_0(\lv)
$$
and observe that $M_-$, $\tv_0(\lv)$ both determine $L_0$-modules. Following Lemma
\ref{lemma2}, we have
\begin{lemma} \label{lemmaA}
	Let $v_+$ be a maximal weight vector for $L$ in $\tv(\lv)$. Then
	$$
	\langle v_+,W\rangle \neq (0).
	$$
$\blacksquare$ \medskip
\end{lemma}
However direct application of this lemma is not quite straightforward. We first observe
that 
$$
\{ M_-,M_-\} = L_{-2} \ (=K_{-2}),
$$
$$
[M_-,L_{-2}] = L_{-2},L_{-2}] = (0).
$$
Using this we may write
\begin{equation}
	U(M_-) = \wedge M_- + L_{-2} U(M_-)
	\label{starappB}
\end{equation}
where 
$$
\wedge M_- \equiv \mathbb{C}\oplus M_-\oplus (M_-\wedge M_-) \oplus\cdots\oplus
\underbrace{(M_-\wedge M_-\wedge \cdots\wedge M_-)}_{\text{$k$ times}}
$$
where 
$$
M_-\wedge M_- \wedge\cdots\wedge M_- \equiv \wedge^i M,\ \ 1\leq i\leq k
$$
is spanned by all generator products
$$
\sigma^0_{\ \mu_1}\sigma^0_{\ \mu_2}\cdots\sigma^0_{\ \mu_i},\ \ 1\leq
\mu_1<\mu_2<\cdots<\mu_i\leq k.
$$
By definition we observe that the $\sigma^0_{\ \mu}$ {\em anti-commute} modulo $L_{-2}$.

Now utilising equation (\ref{starappB}) we write
\begin{align*}
	W &= U(M_-)\tv_0(\lv) \\
	  &= W_- + L_{-2}W
\end{align*}
where
$$
W_- = (\wedge M_-)\tv_0(\lv).
$$
Then we have the following strengthened version of Lemma \ref{lemmaA}:
\begin{lemma}
	Let $v_+$ be an $L$-maximal weight vector in $\tv(\lv)$. Then
	$$
\langle v_+,W_-\rangle \neq 0.
	$$
\end{lemma}
\proof{
	Clearly 
	$$
	\langle v_+,L_{-2}W\rangle = \langle L_2v_+,W\rangle =
	(0).
	$$ 
	Thus
	$$
\langle v_+,W_-\rangle = 0
	$$
	\begin{align*}
\Rightarrow \ \		\langle v_+,W\rangle &= \langle v_+,W_- + L_{-2}W\rangle \\
								 &= (0)
	\end{align*}
	and the result follows from Lemma \ref{lemmaA}.
}

Now we observe that $W_-$ is a completely reducible $L_0$-module and decomposes into
irreducible $L_0$-submodules with highest weights precisely of the form of the branching
theorem, Theorem \ref{branchingtheorem}, with each such module occurring at most once.

The result of Theorem \ref{branchingtheorem} also holds for the low order cases which we
summarise below.

\vspace{.5cm}

\noindent
\underline{{\bf $osp(3|n))\supset osp(2|n)=L$}}

\vspace{.5cm}

Irreducible $L$-modules in irreducible $osp(3|n)$ module $\tv(\lv)$
$$
\lv = \ell \varepsilon + \sum_{\mu=1}^k \lv_\mu \delta_\mu,\ \ \ell\in\frac12 \mathbb{Z}^+
$$
have highest weights of the form
$$
\Lambda = m\varepsilon + \sum_{\mu=1}^k\Lambda_\mu\delta_\mu
$$
satisfying betweenness conditions
$$
\ell\geq m\geq -\ell,\ \ \ell-m\in\mathbb{Z}_+,\ \ \lv_\mu\geq\Lambda_\mu\geq \lv_\mu -1,\
\ 1\leq \mu\leq k,
$$
with each occurring at most once.

\vspace{.5cm}

\noindent
\underline{{\bf $osp(2|n))\supset osp(1|n)=L$}}

\vspace{.5cm}

As above except now $\ell$ can be {\em any} complex number and the allowed $L$-highest
weights are of the form
$$
\Lambda = \sum_{\mu=1}^k\Lambda_\mu\delta_\mu,\ \ \lv_\mu\geq\Lambda_\mu\geq\lv_\mu -1,\
\ 1\leq \mu\leq k,
$$
with each occurring at most once.

\vspace{.5cm}

\noindent
\underline{{\bf $osp(1|n))\supset sp(n)$}}

\vspace{.5cm}

As above except now
$$
\lv = \sum_{\mu=1}^k\lv_\mu\delta_\mu,\ \ \Lambda = \sum_{\mu=1}^k\Lambda_\mu\delta_\mu,\
\ \lv_\mu\geq\Lambda_\mu\geq \lv_\mu-1.
$$


\section*{Appendix C: Comparison to distinguished root system, $m= 2$} 

The branching condition of Section \ref{branchingcondition} (c.f. Appendix B) extends to the cases
$m=1,2$ except that in the former case the canonical subalgebra of $osp(1|n)$ is actually the even
subalgebra $L_0=sp(n)$. It also applies to the embedding $osp(3|n)\supset osp(2|n)$. However, in the
case of the subalgebra $L=osp(2|n) = C(k+1)$ (recall $n=2k$) it is necessary to adopt the
$\mathbb{Z}$-gradation of the other algebras $osp(m|n)$, $m\neq 2$ (see equation (\ref{starp17})).

Since the case $m=2$, (that is, $C(k+1)=osp(2|n)$) is of independent interest, it is worth giving further discussion of this
case which is somewhat unusual in some respects. Now in this case our $o(m)=o(2)$ subalgebra is
1-dimensional and spanned by the Racah generator
$$
\Omega = \sigma^{i=1}_{\ i=2} = \sigma^1_{\ \overline{1}} = -\sigma^2_{\ 1}
$$
(since $\sigma^1_{\ 1} = \sigma^2_{\ 2}=0)$. Our previous choice of $\mathbb{Z}$-grading is
equivalent to choosing the following odd positive roots:
$$
\Phi_1^+ = \{ \pm\varepsilon + \delta_\mu\ |\ 1\leq \mu\leq k\}.
$$
The corresponding odd generators are given by
\begin{equation}
	\sigma^\mu_{\ i},\  \ 1\leq \mu\leq k, \ i=1,2, \label{starp18II}
\end{equation}
and we have the usual $\mathbb{Z}$-grading
$$
L = osp(2|n) = L_{-2}\oplus L_{-1}\oplus L_0 \oplus L_1 \oplus L_2
$$
where $L_0 = o(2)\oplus gl(k)$, $L_{\pm 2}$ are as before and $L_1$ is spanned by generators
(\ref{starp18II}), with a similar result for $L_{-1} = (L_1)^\ddagger$. In this case $L_1$ is a
direct sum of two irreducible $L_0$-modules corresponding to $i=1,2$ in equation (\ref{starp18II}).  

The above is in fact equivalent to making the following choice of simple roots:
$$
\pm \varepsilon+\delta_k,\delta_1-\delta_2,\ldots,\delta_{k-1}-\delta_k.
$$
Hence with this choice we have two {\em odd} simple roots (as distinct from the standard {\em
distinguished} choice of Kac \cite{Kac1977,Kac1978}). Provided this is done our results above on
characteristic identities, branching conditions, reduced matrix elements, etc. all apply to the
subalgebra chain
$$
osp(m|n)\supset osp(m-1|n)\supset\cdots\supset osp(2|n)\supset osp(1|n)\supset osp(0|n)
$$
where we set $osp(0|n)\equiv sp(n)$, with the appropriate identifications of characteristic roots.


To highlight the differences with the distinguished choice of simple roots usually made in
discussing $osp(2|n)$, it is convenient to work with Cartan-Weyl generators $S^a_{\ b}$. In this
notation our single $o(2)$ generator is
$$
\Omega = S^{i=1}_{\ 1} = -S^2_{\ 2},
$$
where now 
$$
S^{i=1}_{\ 2} = S^{i=2}_{\ 1} = 0
$$
identically. With the distinguished set of simple roots we actually have a type I
$\mathbb{Z}$-grading
$$
L = L_{-1}\oplus L_0\oplus L_1
$$
where $L_0=o(2)\oplus sp(n)$ is the even subalgebra and
\begin{align*}
	L_1 &= \mbox{span}\left\{ S^{i=1}_{\ \mu}\ |\ 1\leq \mu\leq n \right\} \\
	L_{-1} &= \mbox{span}\left\{ S^{\mu}_{\ i=1}\ |\ 1\leq \mu\leq n \right\} 
\end{align*}
corresponding to eigenvectors of our $o(2)$ generator $\Omega$, under the adjoint action, with
eigenvalues $\pm 1$ respectively. Thus in this case $\Omega$ is actually a {\em level operator} for
$L = osp(2|n) = C(k+1)$.

This is equivalent to the following choice for our simple roots
$$
\alpha_s = \varepsilon - \delta_1,\ \alpha_1 = \delta_1 -
\delta_2,\ldots,\alpha_{k-1}=\delta_{k-1}-\delta_k,\ \alpha_k = 2\delta_k.
$$
This is the {\em distinguished} choice with a single odd simple root $\alpha_s$ and where $\alpha_i$
($1\leq i\leq k$) are the usual simple roots for $sp(n)$. The corresponding {\em positive} roots are
$$
\Phi^+ = \Phi^+_0 \cup \Phi^+_1
$$
with 
$$
\Phi^+_0 = \{ \delta_\mu \pm \delta_\nu\ |\ 1\leq \mu < \nu \leq k\} \cup \{2\delta_\mu\ |\ 1\leq
\mu\leq k\}
$$
the usual positive roots for $sp(n)$ and
$$
\Phi^+_1 = \{ \varepsilon\pm\delta_\mu\ |\ 1\leq \mu\leq k\}
$$
the odd positive roots.

This is quite different to the choice made above. In this case it is worth noting that
\begin{align*}
	\rho_0 &= \frac12\sum_{\alpha\in\Phi^+_0} \alpha  = \frac12\sum_{\mu = 1}^k(n-2\mu+2)\delta_\mu,\\
	\rho_1 &= \frac12\sum_{\beta\in\Phi^+_1}\beta = \frac{n}{2}\varepsilon,
\end{align*}
$$
\Rightarrow\ \ \rho =  \frac12\sum_{\mu = 1}^k(n-2\mu+2)\delta_\mu  - \frac{n}{2}\varepsilon
$$
which is quite different to the expression determined in equation (\ref{starp20}).


As an example, to highlight the difference between the above two conventions for $L=osp(2|n)$ it is useful to
consider the vector module $V$, which is $(n+2)$-dimensional with even basis vectors $\{ e_i\ |\
i=1,2\}$ and odd basis vectors $\{ e_\mu\ |\ 1\leq\mu\leq n\}$.

\noindent (i) With the distinguished choice, $V$ admits the following $\mathbb{Z}$-gradation into
irreducible $L_0=o(2)\oplus sp(n)$ modules:
$$
V = V_0(\varepsilon)\oplus V_0(\delta_1)\oplus V_0(-\varepsilon)
$$
with $V_0(\pm\varepsilon)$ being 1-dimensional irreducible $L_0$-modules with $o(2)$ highest weights
$\pm \varepsilon$ and $V_0(\delta_1)$ the vector module for $sp(n)$.

\noindent (ii) With our non-distinguished choice on the other hand we have the $\mathbb{Z}$-gradation
$$
V = V_0(\delta_1)\oplus \left[ V_0(\varepsilon)\oplus V_0(-\varepsilon) \right]\oplus V_0(-\delta_k)
$$
with $V_0(\delta_1)$ the vector module of $gl(k)$, $V_0(-\delta_k)$ the dual vector module (minimal
$\mathbb{Z}$-graded component) and $V_0(\varepsilon)\oplus V_0(-\varepsilon)$ the direct sum of two
irreducible $L_0 = o(2)\oplus gl(k)$ modules of dimension 1.

\noindent
{\em Remark}: Whichever choice is made the eigenvalues
$$
\chi_{\Lambda}(C_L) = (\Lambda,\Lambda+2\rho)
$$
must be the {\em same}.


\section*{Appendix D: Explicit characteristic roots for $m\leq3$} 

To illustrate that our formalism also extends to the case $m\leq 2$ we explicitly give the
characteristic roots for these lower values of $m$, after first outlining the case $m=3$ for
contrast.

\noindent
\underline{{\bf $osp(3|n)$}}

Here out highest weights are of the form
$$
\Lambda = \ell\varepsilon + \sum_{\mu=1}^k\Lambda_\mu\delta_\mu,\ \ \ell\in\frac12 \mathbb{Z}_+
$$
and our characteristic roots are
\begin{align*}
	\alpha_{i=1} &= \ell + 3-i-1-\frac{n}{2}=\ell + 1-\frac{n}{2}, \\
\alpha_{\overline{i}} &= \alpha_3 = -\ell - \frac{n}{2},\\
	\alpha_{n+1} &= \alpha_2 = \frac12(3-n-2) = \frac12(1-n)
\end{align*}
with odd roots 
\begin{align*}
	\alpha_\mu &= -\Lambda_\mu + \mu - 1 + (m=3) - n = -\Lambda_\mu + \mu-n+2,\\
	\alpha_{\overline{\mu}} &= \Lambda_\mu - \mu + 1,\ \ 1\leq \mu\leq k.
\end{align*}

\noindent
\underline{{\bf $osp(2|n)$}}

Now our highest weights are of the form
$$
\Lambda = \lambda\varepsilon + \sum_{\mu=1}^k \Lambda_\mu \delta_\mu,\ \ \lambda\in
\mathbb{R}.
$$
Our previous formulae for $\rho_0$, $\rho_1$ and $\rho = \rho_0-\rho_1$ hold exactly as
before by setting $m=2$. Thus our previous formulae apply with $m=2$ to give the
even characteristic roots
\begin{align*}
	\alpha_{i=1} &= \lambda+2-i-1-\frac{n}{2} = \lambda - \frac{n}{2},\\
	\alpha_{\overline{i}} &= \alpha_{i=2} = -\lambda + i-1-\frac{n}{2} = -\lambda -
	\frac{n}{2}
\end{align*}
and odd characteristic roots
\begin{align*}
	\alpha_\mu &= -\Lambda_\mu + \mu + m-n-1 = -\Lambda_\mu + \mu - n+1,\\
	\alpha_{\overline{\mu}} &= \Lambda_\mu - \mu + 1,\ \ 1\leq\mu\leq k.
\end{align*}

\noindent
\underline{{\bf $osp(1|n)$}}

Now our highest weights are of the form
$$
\Lambda = \sum_{\mu=1}^k\Lambda_\mu\delta_\mu.
$$
In this case $\rho_0$ and $\rho_1$ (and thus $\rho$) are formally exactly as before by
simply setting $m=1$:
$$
\rho_0 = \frac12\sum_{\mu=1}^k (n-2\mu+2)\delta_\mu,\ \ \rho_1 =
\frac12\sum_{\mu=1}^k\delta_\mu.
$$
This gives the single even characteristic root (corresponding to $i=h+1 = 1$ since $m=1$
$\Rightarrow$ $h=0$ for odd $m=1$)
$$
\alpha_i = h+1 = \frac12(m-n-2) = -\frac12(n+1)
$$
together with the odd characteristic roots
\begin{align*}
	\alpha_\mu &= -\Lambda_\mu + \mu -1 + m-n = -\Lambda + \mu -n,\\
\alpha_{\overline{\mu}} &= \Lambda_\mu - \mu + 1,\ \ 1\leq\mu\leq k.
\end{align*}

\noindent
\underline{{\bf $sp(n) = osp(0|n)$}}

Here we again have
$$
\Lambda = \sum_{\mu=1}^k\Lambda_\mu\delta_\mu,
$$
but now our algebra corresponds to the even Lie algebra $sp(n)$. In this case we have
$$
\rho_1 = 0,\ \ \rho_0 = \frac12\sum_{\mu=1}^k (n-2\mu + 2)\delta_\mu 
$$
which is the same as before if we formally set $m=h=0$. In this case there are no {\em
even} characteristic roots and our {\em odd } characteristic roots are
\begin{align*}
	\alpha_\mu &= -\Lambda_\mu + \mu -1+m-n = -\Lambda_\mu + \mu -1 -n,\\
	\alpha_{\overline{\mu}} &= \Lambda_\mu - \mu +1
\end{align*}
which are given by our previous formula by setting $m=0$.


%
%
%


\end{document}